\documentclass[useAMS,usenatbib,usegraphicx]{mn2e}

\title[$N$-body model of M67]{A complete $N$-body model of the old open cluster M67}
\author[J. R. Hurley et al.]{Jarrod R. Hurley$^{1,2}$\thanks{
E-mail: jarrod.hurley@sci.monash.edu.au (JRH)}, 
Onno R. Pols$^{3}$, 
Sverre J. Aarseth$^{4}$ and Christopher A. Tout$^{4}$\\
$^{1}$Centre for Stellar and Planetary Astrophysics, School of Mathematical Sciences,
            Monash University, VIC 3800, Australia\\
$^{2}$Department of Astrophysics, American Museum of Natural History, 
            Central Park West at 79th Street, New York, NY 10024, USA\\
$^{3}$Astronomical Institute, Utrecht University, Postbus 80000, 3508 TA Utrecht, the Netherlands\\
$^{4}$Institute of Astronomy, Madingley Road, Cambridge CB3 0HA, UK}

\begin{document}

\date{Accepted 2005 Month xx. Received 2005 Month xx; in original form 2005 May 26} 

\pagerange{\pageref{firstpage}--\pageref{lastpage}} \pubyear{2005}

\maketitle

\label{firstpage}

\begin{abstract}
The old open cluster M67 is an ideal testbed for current cluster evolution
models because of its dynamically evolved structure and rich stellar
populations that show clear signs of interaction between stellar, binary and
cluster evolution.  Here we present the first truly direct $N$-body model
for M67, evolved from zero age to $4\,$Gyr taking full account of cluster
dynamics as well as stellar and binary evolution. Our preferred model
starts with $36\,000$ stars ($12\,000$ single stars and $12\,000$ binaries) and a
total mass of nearly $19\,000 M_\odot$, placed in a Galactic tidal field at
$8.0\,$kpc from the Galactic Centre.  Our choices for the initial conditions
and for the primordial binary population are explained in detail.  At
$4\,$Gyr, the age of M67, the total mass has reduced to $2\,000 M_\odot$
as a result of mass loss and stellar escapes. The mass and half-mass radius
of luminous stars in the cluster are a good match to observations although the
model is more centrally concentrated than observations indicate. The
stellar mass and luminosity functions are significantly flattened by
preferential escape of low-mass stars. We find that M67 is dynamically old
enough that information about the initial mass function is lost, both from
the current luminosity function and from the current mass fraction in white
dwarfs.

The model contains 20 blue stragglers at $4\,$Gyr which is slightly less
than the 28 observed in M67. Nine are in binaries. The blue stragglers 
were formed by a variety of means and we find formation paths for the
whole variety observed in M67.  Both the primordial binary
population and the dynamical cluster environment play an essential role in
shaping the population.  A substantial population of short-period 
primordial binaries (with periods less than a few days) 
is needed to explain the observed number of blue stragglers in M67. 
The evolution and properties of two thirds of the blue stragglers, 
including all found in binaries, 
have been altered by cluster dynamics and nearly half would not have
formed at all outside the cluster environment.  On the other hand, the
cluster environment is also instrumental in destroying potential BSs from
the primordial binary population, 
so that the total number is in fact
slightly smaller than what would be expected from evolving the same binary
stars in isolation.

\end{abstract}


\section{Introduction}
\label{s:intro}

Star clusters have long been recognized as important tools for 
understanding many astrophysical processes. 
As such they are the focus of many observational programs, 
both from the ground (e.g. Kalirai et al. 2003; Kafka et al. 2004) 
and space (e.g. Grindlay et al. 2001; Piotto et al. 2002; Richer et al. 2004). 
Dynamical modelling had its beginnings over four decades ago 
(von Hoerner 1960; Aarseth 1966; van Albada 1968) 
but it is only recently that the models have reached a state where genuine 
comparison with cluster observations is possible. 
This is the result of software advances that have improved realism and 
speed as well as a huge increase in hardware performance. 
The realm of globular clusters is still out of reach of realistic direct $N$-body 
models. 
So it is open clusters that garner initial attention in the attempt 
to confront cluster models with observations and vice-versa. 
In particular, old open clusters are of interest because these are dynamically 
well-evolved and offer the chance to observe the effects of interaction 
between stellar, binary and cluster evolution. 
A good example is M67 which contains an abnormally large number of 
blue straggler (BS) stars for its age and size (Ahumada \& Lapasset 1995), 
as well as a good proportion of X-ray sources (van den Berg et al. 2004). 
Both are indicators that the cluster environment has affected the evolution 
of the stars. 
M67 is also close enough to be well studied and within range of 
current $N$-body modelling capabilities\footnote{M67 has been 
identified by the MODEST collaboration (Sills et al. 2003) as an ideal 
cluster for various groups working on cluster evolution models to 
compare simulation techniques and use the observed data to calibrate 
the models: see {\tt http://manybody.org/modest/projects.html}.}. 
The idea is that by matching the characteristics of a model cluster and 
the observed cluster we can learn about the initial conditions, binary 
population and dynamical evolution of the cluster. 
In doing this we must bear in mind the need to marry theoretical and 
observational data reduction techniques (Portegies Zwart et al. 2004). 

The interest in M67 and its BSs stems not only from the large number 
of BSs but also from their diverse nature. 
Observations indicate that 60 per cent are in spectroscopic 
binaries (Milone \& Latham 1992; Latham \& Milone 1996). 
Of those in binaries the orbits are a mixture of short-period and 
eccentric, long-period and eccentric, and long-period and circular. 
Blue stragglers are identified in a cluster colour-magnitude diagram (CMD) 
by their position above and blueward of the main-sequence turn-off. 
A popular explanation for their existence is that they are rejuvenated 
main-sequence (MS) stars that have gained hydrogen-rich material. 
Mass transfer in a binary from a companion to a MS star is an obvious 
means by which this can occur. 
There are three main scenarios (Kippenhahn, Wiegert \& Hoffmeister 1967) 
distinguished by the stellar type of the mass donor, MS star (Case~A), 
sub-giant or red giant (Case~B), 
or asymptotic giant branch (AGB) star (Case~C). 
Case~A mass transfer can produce a BS in a very short-period Algol 
system, but in many cases ends in coalescence of the two MS stars when 
angular momentum is lost from the system and the orbit shrinks. 
This produces single BSs. 
Case~B mass-transfer occurs in slightly wider binaries and results in a 
short-period circular binary containing a BS with a white dwarf companion. 
BSs in longer period binaries may be produced by Case~C mass transfer 
or by accretion of material from an AGB star wind. 
The binary orbit will be circular except in some cases of wind accretion. 
However current models of binary tides and wind accretion only allow 
for accretion to occur and the orbit to retain an eccentricity in binaries 
with periods in excess of a few thousand days and the degree of accretion 
drops for wider orbits. 
This is a known problem in attempts to explain the existence of Barium 
stars in eccentric binaries (Pols et al. 2003). 
Thus binary evolution alone fails to explain the full range of BSs found 
in M67 and, in particular, the short-period eccentric binaries.
It also fails to explain the so-called super-BS observed in M67 (Leonard 1996) 
which has a mass greater than twice that of the cluster MS turn-off mass, 
$M_{\rm TO}$. 
Blue stragglers could also form as a result of direct collisions between 
MS stars although this is unlikely as the timescale for such an event 
to occur in an open cluster is too long 
(Press \& Teukolsky 1977; Mardling \& Aarseth 2001). 

In Hurley et al. (2001) we presented a semi-direct $N$-body model of M67. 
We showed that the combination of the cluster environment and 
a large population of binaries is able to generate formation paths for 
all of the BSs observed in M67. 
BSs were found in wide eccentric binaries because the BSs were exchanged 
into such orbits in three- and four-body encounters subsequent to their 
formation. 
Cases where two MS stars forming the inner binary of a triple system 
merged and remained bound to the third star were also observed and 
these produced BSs in short-period eccentric orbits. 
Roughly half of the BSs formed during the simulation were the result of 
standard binary evolution in primordial binaries. 
The remainder owed their existence in some way to dynamical encounters. 
Some were the result of perturbations to primordial binaries. 
The orbits of these binaries are dynamically altered so that mass-transfer 
begins whereas the stars would have remained detached if evolved in isolation. 
Collisions between MS stars at periastron in highly eccentric binaries also 
proved to be an efficient BS formation pathway. 
These eccentric binaries were the result of exchange interactions or 
perturbations to the orbits of primordial binaries. 
Super-BSs were also formed after the merger of three or more MS stars 
-- an example of this sees a single BS formed via Case~A mass transfer 
and then exchanged into an eccentric binary where it collides with its new 
MS companion. 
So this simulation was successful in showing that the cluster environment is 
able to boost BS production and that the diverse nature of the M67 
BSs can be explained if they formed by a variety of mechanisms. 
However, the model did not generate enough BSs in binaries at any single 
moment -- at most 25\% of the BSs were in binaries when the model was 
at or near the age of M67. 

The main failing of the Hurley et al. (2001) M67 model was that it is semi-direct. 
By this we mean that, owing to computational constraints, the model was 
not evolved by the direct $N$-body method for its entirety. 
At the time special-purpose hardware for computing gravitational forces was 
available in the form of the GRAPE-4 (Makino \& Taiji 1998) and while this had 
given a substantial increase in the performance of $N$-body codes it 
still did not allow a complete model of M67 to be evolved from birth in a timely 
manner. 
The complication that M67 has a high frequency of binaries further 
reduced the simulation rate.  
The method employed by Hurley et al. (2001) was to take a population 
of $5\,000$ single stars and $5\,000$ binaries and evolve these from 
birth to an age of $2.5\,$Gyr according to a rapid stellar and binary evolution 
algorithm (Hurley, Tout \& Pols 2002) -- standard population synthesis. 
This evolved population was then used as input to the $N$-body code 
and evolved further to the age of M67 (about $4\,$Gyr, see next section). 
The results of previous $N$-body simulations were used as a guide when 
selecting the stars and binaries for the model and to generate the density 
profile for the starting $N$-body model. 
So every effort was made to account for the effect of processes such as 
mass-segregation during the $2.5\,$Gyr of cluster evolution that was skipped. 
However the situation was far from ideal. 
Since that time 
$N$-body capabilities have improved markedly with the arrival of the 
next generation of special-purpose hardware, the GRAPE-6 (Makino et al. 2003), 
and its teraflops speed. 
In the meantime there have 
also been software and host-CPU speed-ups to aid with binaries 
(see Aarseth 2003). 
As a result binary-rich open clusters are now within reach of direct $N$-body 
codes (Hurley \& Shara 2003) as are 
larger single-star models (Baumgardt \& Makino 2003). 
So we can now perform a direct model of M67. 

In this paper we 
attempt to create a direct $N$-body model of the old open cluster M67. 
We discuss the model in comparison with observations of M67, 
looking at the overall structural properties and the nature 
of the stellar populations. 
In Section~\ref{s:initial} we describe the parameters of our starting model 
and in Section~\ref{s:binary} we look at the nature of the primordial 
binary population in some detail. 
We describe our simulation method in Section~\ref{s:method}. 
Section~\ref{s:result1} contains the results of our first attempt at a model 
of M67 and then in Section~\ref{s:result2} we present our favoured M67 model. 
This is followed by a 
discussion and summary of our results in Section~\ref{s:discus}.

\section{The initial model}
\label{s:initial}

The goal is to evolve an $N$-body model cluster from zero-age to obtain 
a model that resembles M67 at its current age in terms of both structural 
parameters and stellar populations. 
Unfortunately we cannot observe M67 at zero-age so in setting up our 
initial model we must be guided by observations of M67 at present in 
combination with the behaviour of previous cluster models and also 
observations of young open clusters. 
In this section we discuss in turn our choices for the metallicity and
age of M67, the binary fraction, the stellar initial mass function,
the initial mass of the cluster, the tidal field and the density
profile. 
The details of the primordial binary population are left for
Section~\ref{s:binary}. 
We consider two models (blandly called Model~1 and Model~2) 
which differ only in the initial mass and the tidal field, and in the 
period distribution of the primordial binaries. 
All other choices are the same for both models. 

It is generally accepted that M67 stars are of solar composition 
(Hobbs \& Thorburn 1991; Tautvai\u{s}ien\.{e} et al. 2000). 
There is no reason to believe that the metallicity of M67 has varied 
over its life so we take $Z = 0.02$ for our initial model. 
The age of M67 is less settled with values quoted in the literature 
ranging from $3.2 \pm 0.4\,$Gyr (Bonatto \& Bica 2003) to as high as 
$6.0\,$Gyr (Janes \& Phelps 1994). 
In Hurley et al. (2001) we assumed an age of $4.2\,$Gyr based on 
fitting the isochrones of Pols et al. (1998) to the available data. 
Recently VandenBerg \& Stetson (2004) derived an age of $4.0\,$Gyr 
for M67. 
We shall investigate behaviour around $4\,$Gyr in the current study while 
bearing in mind that the cluster may be slightly younger or older. 

\begin{figure}
\includegraphics[width=84mm]{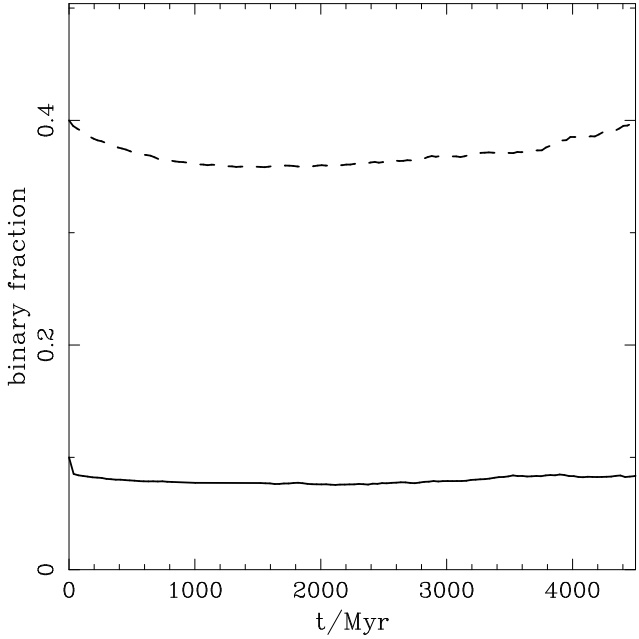}
\caption{
Evolution of binary fraction with age for an $N$-body cluster model 
starting with $18\,000$ stars and $2\,000$ binaries 
(Shara \& Hurley 2002: solid line). 
Also shown is a model that started with $12\,000$ stars and $8\,000$ 
binaries (Hurley \& Shara 2003: dashed line). 
\label{f:fig1}}
\end{figure}

M67 is definitely a binary-rich cluster. 
Montgomery, Marschall \& Janes (1993) found that at least 38 per cent of the 
cluster stars are binaries based on analysis of the 
photometric binary sequence in the colour-magnitude diagram (CMD). 
A high binary frequency was confirmed by Fan et al. (1996) and 
Richer et al. (1998) who both found 50 per cent to be consistent with their data. 
How this translates to a primordial binary fraction depends somewhat 
on the parameters of the primordial binaries and the evolution 
of the cluster. 
A good indicator is to look at the evolution of binary fraction in previous 
open cluster models. 
Shara \& Hurley (2002) performed simulations starting with $18\,000$ 
single stars and $2\,000$ binaries -- a primordial binary fraction of 0.1. 
The orbital separations of the binaries were drawn from the  
distribution suggested by Eggleton, Fitchett \& Tout (1989) with a peak 
at $30\,$au and limits of $0.03\,$au (about $6 R_\odot$) and $30\,000\,$au. 
This distribution was based on the Bright Star Catalogue (Hoffleit 1983) and 
is in agreement with the period distribution found by 
Duquennoy \& Mayor (1991). 
After $4.5\,$Gyr of cluster evolution, Shara \& Hurley (2002) found that the 
binary fraction was still close to 0.1. 
The evolution of binary fraction is shown in Figure~\ref{f:fig1} for one of 
their models. 
Also shown is the binary fraction for a simulation presented by Hurley \& Shara (2003) 
which began with $12\,000$ single stars and $8\,000$ binaries. 
They used the same distribution for the initial orbital separations but set the 
maximum at $200\,$au. 
This means the 40 per cent binaries would actually represent a 
population of 55 per cent binaries if the upper limit were set at $30\,000\,$au. 
In both cases the binary fraction remains similar as the cluster evolves. 
This can be understood because, while encounters may destroy soft binaries and 
a combination of dynamical hardening and binary evolution may destroy 
very hard binaries, the preferential escape of low-mass single stars from the 
cluster via tidal stripping counteracts these effects on the binary fraction. 
With a high fraction of primordial binaries residing in the high density core we 
do expect many binaries to be lost in binary-binary encounters. 
However, even in this region of the cluster the binary 
fraction does not drop because single stars are ejected from the core in the 
same encounters (Aarseth 1996). 
The conclusion is that, for reasonable choices of the binary parameters, experience 
shows that in open cluster simulations the primordial binary fraction is well 
preserved as the cluster evolves. 
Therefore we assume a primordial binary fraction of 0.5 and elaborate on the 
particulars of these binaries in the next section. 

Single star initial masses are chosen from the initial mass function (IMF) of 
Kroupa, Tout \& Gilmore (1993, KTG93) between the mass limits of 0.1 and 
$50 M_\odot$. 
All stars are assumed to be on the zero-age main sequence (ZAMS) when 
the simulation begins. 
This means that we neglect any age spread in the 
cluster stars that may be caused by differences in the pre-main sequence 
evolution timescales or by more than one epoch of star formation. 
We also assume that any residual gas from the star formation 
process has been removed. 

The number of stars in the initial model, or alternatively the initial mass of 
the cluster $M_0$, is an important quantity. 
To estimate it is difficult for us because determination of the current mass of M67 from 
observations suffers from incompleteness and converting this to an initial 
mass requires knowledge of the mass-loss history from the cluster. 
A lower limit of approximately $800 M_\odot$ for the current mass of M67 was 
set by Montgomery et al. (1993). 
Later Fan et al. (1996) established that M67 has about $1\,000 M_\odot$ in 
nuclear-burning stars with masses greater than $0.5 M_\odot$. 
We shall refer to this as the luminous mass, $M_{\rm L}$. 
Hurley et al. (2001) showed that this corresponds to a total cluster mass, $M$, 
of $2\,500 M_\odot$ and $M_0 \simeq 3\,500 M_\odot$ if stellar and binary evolution 
are taken into account but the influence of the dynamical cluster environment 
is ignored.
However, stars in a cluster are subject to mass segregation and this 
process, combined with the stripping of stars by the tidal field of the 
Galaxy, alters the mass function of the cluster over time to give a deficiency 
of low-mass stars. 
Looking at previous $N$-body data available to us 
(Shara \& Hurley 2002; Hurley \& Shara 2003)
we find that $M_{\rm L} = 1\,000 M_\odot$ is more likely to represent a 
total mass of approximately $1\,400 M_\odot$ after $4\,$Gyr of stellar, 
binary and cluster evolution. 
We note that the limiting radius of the Fan et al. (1996) observations 
was $10\,$pc while the tidal radius of M67 is greater 
and therefore the actual mass of the cluster may be slightly higher. 

Next we must convert the current mass to an initial mass. 
Hurley et al. (2001) estimated that a starting model of $40\,000\,$stars 
would be required to make a full model of M67. 
This was based on models with an 
escape rate parameter $k_{\rm e} = 0.3$ (see their equation 18). 
They also took the initial half-mass radius to be $1\,$pc which gave 
a starting model well within its tidal radius. 
$N$-body models starting with $30\,000\,$stars and no primordial 
binaries have recently been presented by Hurley et al. (2004). 
Analysis of these models shows that $k_{\rm e} = 0.3$ is once again 
suitable to describe the average escape rate over the lifetime of the cluster. 
Using this value, and taking $M_0 = 15\,000 M_\odot$ for a cluster that 
fills its tidal radius initially and has a half-mass radius, $r_{\rm h}$, 
of $5\,$pc (giving a half-mass relaxation timescale of about $300\,$Myr), 
we find that approximately $2\,000 M_\odot$ in stars is left after $4\,$Gyr. 
This neglects mass loss owing to stellar evolution and there is also a 
degree of redundancy in the analysis but it shows that something of 
this order is required for the starting model. 
The $30\,000\,$single-star models of Hurley et al. (2004) actually 
had $M \approx 5\,800 M_\odot$ at $4\,$Gyr which is too high for M67. 
However, it has been shown (Hurley \& Shara 2003) 
that the presence of a large number of primordial 
binaries increases the escape rate of stars and this is confirmed in 
our future paper on the general trends of binary-rich open cluster evolution 
(Hurley et al., in preparation). 
We note that the escape rate analysis of Hurley et al. (2001) does not 
consider the effect of binaries. 
Based on all this we start in our first attempt (Model~1) 
with $9\,000$ single stars and $9\,000$ binaries 
which gives $M_0 \simeq 14\,400 M_\odot$, comparable to that of the 
$30\,000\,$ single-star models. 

In Model~1 we assume for simplicity that the cluster 
is subject to a standard Galactic tidal field for the Solar neighborhood, 
a circular orbit with a speed of $220\, {\rm km} \, {\rm s}^{-1}$ at a distance 
of $8.5\,$kpc from the Galactic centre. 
For a mass of $14\,400 M_\odot$ this gives a tidal radius, $r_{\rm t}$, of 
$34\,$pc. 
M67 actually has a slightly eccentric orbit with a perigalacticon of $6.8\,$kpc 
and an apogalacticon of $9.1\,$kpc (Carraro \& Chiosi 1994). 
We will therefore consider a second model (Model~2) starting with 
$12\,000$ single stars and $12\,000$ binaries on an orbit at $8.0\,$kpc 
from the Galactic centre but leave further details of this to Section~\ref{s:result2}, 
after we have presented the results for the first model. 

The density profile chosen for our starting model is that of a Plummer model 
(Plummer 1911;  Aarseth, H\'{e}non \& Wielen 1974). 
There is no reason to believe that a King model (King 1966), the major 
alternative, should describe the initial state of an open cluster as this set 
of models was originally conceived to describe dynamically evolved 
globular clusters. 
Kroupa, Aarseth \& Hurley (2001) built a model of the Orion Nebula cluster 
which for simplicity used a Plummer density profile. 
They also assumed that the model contained twice the mass in gas as it had 
in stars and then after $0.6\,$Myr of $N$-body evolution the gas was removed 
smoothly. 
The result was not only a bound cluster but, after $100\,$Myr, the projected 
radial density profile was a good fit to that of the Pleiades. 
In fact the choice is not likely to be crucial because previous $N$-body studies 
(e.g. Hurley \& Shara 2003) have shown that an initial Plummer model quickly 
evolves to resemble a King profile. 
The density profile in combination with the assumption that the stars are 
in virial equilibrium generates the initial positions and isotropic velocities 
of the stars. 
Formally the Plummer model extends to infinite radius so in 
the rare case of large distance (more than $10 r_{\rm h}$) a rejection is applied. 
The length-scale parameter is chosen such that the cluster fills its tidal 
radius from the beginning of the simulation. 
The position of the outermost star lies at a distance of $34\,$pc from the 
cluster centre. 
The choices for both starting models are summarized in Table~\ref{t:table1}.

\section{Parameters of the primordial binary population}
\label{s:binary}

When setting up a population of primordial binaries we must choose how 
to distribute their orbital parameters. 
For each binary we need an orbital separation, $a$ (or period, $P$), 
an eccentricity, $e$, and masses for the two stars (or one mass, which may  
be the total binary mass, $M_{\rm b}$, and a mass ratio, $q$). 
The distribution function chosen for the orbital separation or period is generally 
guided by observations with popular choices being a flat distribution of 
$\log a$ (Abt 1983) 
or a log-normal distribution (Eggleton, Fitchett \& Tout 1989; Duquennoy \& Mayor 1991). 
For the eccentricity a choice is usually made between a thermal distribution 
(Heggie 1975), $f(e) = 2 e$, or simply to 
have all orbits initially circular (although $e = 0.0$ is generally avoided for 
numerical reasons). 
When determining the masses of the component stars these may be 
uncorrelated, each mass is drawn independently from the same IMF, 
or a distribution of mass ratios may be assumed. 

In view of the relatively large number of blue stragglers found in M67,  
Hurley et al. (2001) performed a binary population synthesis study to 
determine which combination of distribution functions for the orbital parameters 
of the primordial binaries would maximize the number of blue stragglers 
produced. 
Binaries were evolved according to the binary star evolution (BSE, 
Hurley, Tout \& Pols 2002) code. 
They found that choosing the orbital separation from a flat distribution of 
$\log a$ in combination with correlated masses could possibly reproduce 
the M67 BS number provided that all BS producing binaries are retained 
by the cluster over its lifetime. 
So in this case standard evolution of primordial binaries is enough without 
the dynamical interactions between cluster stars. 
However, the nature of the synthetic BS population does not correspond with 
that of the observed population. 
Only 25 per cent of the BSs are in binaries and 
all of these have circular orbits with almost all having an orbital period of 
less than a year. 
Therefore, even if the cluster environment does not have a role to play in 
creating BS formation channels, it must be affecting the evolution of the binaries 
and the nature of the BS population. 

The case of uncorrelated component masses was ruled out by Hurley et al. (2001) 
because it leads to too few BSs by far. 
Their favoured population synthesis model had the Eggleton, Fitchett \& Tout (1989) 
separation distribution with a peak at $10\,$au and a maximum of $200\,$au 
in combination with correlated masses and a thermal eccentricity distribution. 
The $200\,$au limit was based on the realisation that binaries wider than this 
would be weakly bound and broken-up by dynamical encounters. 
This model predicted that about ten of the M67 BSs could be produced from 
standard binary evolution of primordial binaries with 30\% of the BSs in 
binaries, all with circular orbits. 
Hurley et al. (2001) then evolved their semi-direct model 
with the primordial binaries set up in this way. 
When the model reached an age of $4.2\,$Gyr there were 22 BSs present. 
Unfortunately only one of the BSs present at $4.2\,$Gyr was still in a binary. 

A problem with the Hurley et al. (2001) semi-direct M67 model was that no BSs 
were found in wide circular binaries at any time during the evolution while 
two are currently observed in M67 (Latham \& Milone 1996). 
Exchange interactions do not tend to produce binaries in circular orbits so it 
would seem that the evolution of primordial binaries must produce any 
BSs in wide circular binaries. 
This can occur as a result of stable Case~C mass transfer from an AGB star 
or via wind accretion. 
On inspection of the binary population in their starting model in combination 
with the setup of the BSE algorithm used to evolve the binaries, Hurley et al. (2001) 
discovered that no such binaries were expected. 
They found that if they repeated the binary population synthesis with the 
wind velocity reduced by a factor of two\footnote{The wind velocity parameter, 
$\beta$ was reduced from 1/2 to 1/8 with the lower value more in keeping 
with observations} 
and the weaker criterion of Webbink (1988) used to detect the onset of dynamical 
mass transfer (as opposed to the default for the BSE algorithm) that the number of 
BSs predicted at $4.2\,$Gyr rose by 25 per cent for their favoured choice of 
distribution functions. 
More important the mix of the BS population  was 22 per cent in wide circular binaries 
(up from 0 per cent), 22 per cent in close circular binaries (down from 28 per cent) and 
56 per cent single (down from 72 per cent). 
We use a period of $1\,000\,$d to make the distinction between close and wide 
binaries. 
In this paper we have used $\beta = 1/8$ and the Webbink (1988) test for dynamical 
mass transfer in the BSE algorithm at all times, both when conducting population 
synthesis and within the $N$-body code. 

An alternative approach to setting up the orbital parameters of the primordial 
binaries is to use the method of pre-MS tidal evolution developed by Kroupa (1995b). 
Here it is assumed that the majority of low-mass stars form in aggregates that 
dissolve within $10\,$Myr (Kroupa 1995a) and that correlations between orbital 
parameters observed for short-period systems are the result of evolution in young 
binaries (less than $10^5\,$yr) as the two accreting protostars interact. 
A birth period distribution can be constructed following Kroupa (1995b) 
and tidal evolution is imposed on the eccentricity 
to replicate the observed deficiency of eccentric 
orbits for short-period G dwarfs. 
A correlation between the masses of the binary stars is also noted (Kroupa 1995b). 
To investigate how this affects the production of BSs we set up the 
primordial binaries in the following manner. 
First the binary mass is chosen from the IMF of Kroupa, Tout \& Gilmore (1991) 
and the component masses are set by choosing a mass ratio from a uniform 
distribution. 
This is the same method as used by Hurley et al. (2001) when they assumed 
correlated masses for the binary stars. 
Next a period is chosen according to the distribution given by Kroupa 
(1995b, equation 8) but with the generating function given in 
Kroupa (1995a, equation 11b) with $\delta = 2.5$, $\eta = 45$ and 
$\log P_{\rm min} / {\rm d}  = 0$. 
This gives $\log P_{\rm max} / {\rm d} = 8.43$ for the distribution. 
An eccentricity is then chosen from a thermal distribution (Heggie 1975). 
If the periastron distance, $R_{\rm peri}$, for these parameters is less 
than five times the ZAMS 
radius of the primary, because pre-MS stars can typically be of this size 
(Kroupa 1995b), it is assumed that a collision would have occurred 
during the initial evolution and a new set of parameters is chosen.  
When this test is passed the birth eccentricity, $e_{\rm b}$, is modified 
according to tidal evolution by 
\begin{equation}
\ln e_{\rm i} = - \left( \frac{\lambda R_\odot}{R_{\rm peri}} \right)^{\chi} + \ln e_{\rm b} \, 
\label{e:eigen}
\end{equation}
(Kroupa 1995b: equations 3b and 4), where $\lambda = 28$, $\chi = 0.75$ 
and $e_{\rm i}$ is the resulting eccentricity of the primordial binary. 

Generating up $500\,000$ binary stars and evolving them with the BSE algorithm 
we find that only 1.5 BSs per $10\,000$ primordial binaries are predicted 
at $4\,$Gyr. 
This is in contrast to the model of Hurley et al. (2001) which predicts 
11 BSs per $10\,000$ primordial binaries. 
Furthermore all of the tidally modified BSs are in wide circular binaries -- 
no Case~A or Case~B mass transfer occurred. 
Recently Davies, Piotto \& De Angeli (2004) have claimed that primordial 
binaries make mostly BSs in wide binaries and that the Case~A/merger 
scenario is not dominant. 
This is based on the Preston \& Sneden (2000) observations of field 
blue metal-poor (BMP) stars. 
However, from the Preston \& Sneden (2000) data, it seems that if 
we assume that all BMP stars are BSs, then a possible mix of these BSs 
is 40 per cent from Case~A evolution (single BS), 10 per cent from Case~B 
(BS in short-period binary) and 50 per cent from Case~C evolution (BS in 
wide binary). 
Furthermore, about 50 per cent of the wide binaries containing a BMP star 
have eccentric orbits and a possible explanation of these is that the BMP star 
(or BS) formed when the two stars comprising the inner binary of a triple 
merged. 
This would indicate that a large fraction of BMP stars were formed via the 
Case~A/merger path. 
All we say on this for now is that much more could be done to constrain 
parameter distributions using population synthesis 
in combination with data from field stars and open cluster populations. 
We plan to look at this further in the near future. 

We will proceed with the Kroupa (1995b) tidal evolution method, 
as described above, to generate the primordial binary population of 
our Model~1. 
This predicts about one BS from the primordial binaries 
so in effect the cluster environment must do all the work to create the 
observed population. 
Even though we expect that this will not be the case it is of interest 
to see if the dynamics can do it alone. 

In the previous section we mentioned that we will also evolve a 
second model 
which will start with more stars than Model~1 and orbit at a closer 
distance to the Galactic centre 
(see Section~\ref{s:result2} for details). 
For this model we also want the 
primordial binaries to be expected to produce BS numbers 
of the order of what is observed in M67. 
The way we select the masses stays the same, $M_{\rm b}$ from 
Kroupa, Tout \& Gilmore (1991) and $n(q) = 1$. 
However, instead of using the Kroupa (1995b) period distribution we 
simply select the initial separation from a flat distribution of 
$\log a$ and impose an upper cutoff at $50\,$au. 
For a binary with $M_{\rm b} = 1 M_\odot$ this corresponds to a period 
of $1.3 \times 10^5\,$d. 
A population of 50 per cent binaries chosen in this way actually 
represents a population of 100 per cent binaries if 
all periods out to $\log P_{\rm max} / {\rm d} = 8.43$ are allowed, as in Model~1. 
The assumption then is that all binaries born with $a > 50\,$au 
quickly disassociate by weak encounters with other stars. 
Ignoring these non-interacting binaries of low binding energy will not 
affect the evolution of the cluster or the formation of stars such as BSs. 
At the lower end of the separation distribution we reject any  
that is less than twice the sum of the ZAMS radii of the component stars. 
We then choose an eccentricity from a thermal distribution (Heggie 1975), 
as before, and modify it according to the Kroupa (1995b) tidal evolution test 
(Eq.~\ref{e:eigen}) if necessary. 
We note that Kroupa (1995a) claims that a flat distribution of $\log  P$ is 
consistent with pre-MS orbital data 
and that the Duquennoy \& Mayor (1991) survey of binaries in the solar 
neighbourhood with G-dwarf primaries does not rule out a flat distribution 
of $\log a$. 
Evolving $500\,000$ binaries set up in this way with the BSE algorithm 
gives 21 BSs at $4\,$Gyr per $10\,000$ primordial binaries. 
The population synthesis predicts that 75 per cent of these should be single, 
13 per cent in short-period circular binaries, and the remaining 12 per cent  
in wide circular binaries. 
Thus the interest for Model~2 is in seeing if the combination of the cluster 
environment and the evolution of the primordial binaries can put the 
BSs in the required mix of living arrangements, while also maintaining 
the expected number of BSs.

\section{Simulation Method}
\label{s:method}

We use the {\tt NBODY4} code (Aarseth 1999) to model the dynamical 
evolution of star clusters. 
Basic integration of the equations of motion is performed by the Hermite 
scheme (Makino 1991) which employs a fourth-order force polynomial 
and exploits the fast evaluation of the force and its first time derivative 
by the GRAPE-6 (Makino et al. 2003). 
A time-step scheme comprising a series of hierarchical levels (McMillan 1986) 
allows each star to evolve on its own natural dynamical timescale while 
forcing a block of particles to be advanced at each cycle so that 
efficiency does not suffer. 
Also, only one prediction is required for each block step. 
Regularization techniques are used to treat perturbed two-body motion 
in an accurate and efficient manner (Mikkola \& Aarseth 1998) with 
an extension to chain regularization (Mikkola \& Aarseth 1993) to deal  
with compact subsystems of up to six bodies. 
A semi-analytical criterion developed by Mardling \& Aarseth (2001) is 
utilized to detect and evolve stable hierarchical triple and quadruple systems 
which otherwise would prove extremely time consuming by 
direct integration. 
Exchange interactions in encounters between single stars 
and binaries, or binary-binary encounters, where the member of a binary 
is displaced by an incoming star are included in this treatment. 
Direct collisions between stars (Kochanek 1992) and the formation of 
binaries in three- and four-body encounters are also allowed. 
Binary orbits may also become chaotic owing to perturbations from 
nearby stars and this scenario is modelled using the 
Mardling \& Aarseth (2001) algorithm. 

In {\tt NBODY4} stellar and binary evolution are performed in step with the 
dynamical evolution so that interaction between these processes is 
modelled consistently (Hurley et al. 2001). 
Stellar evolution is included in the form of the single star evolution (SSE, 
Hurley, Pols \& Tout 2000) algorithm. 
This is a package of analytical formulae fitted to the detailed models 
of Pols et al. (1998) that covers all phases of evolution from the ZAMS 
up to, and including, remnant phases. 
It is valid for masses in the range $0.1 - 100 M_\odot$ and metallicity 
can be varied. 
The SSE package contains a prescription for mass loss by stellar winds 
that is utilized in {\tt NBODY4}. 
It follows the evolution of rotational angular momentum for each star 
after the ZAMS spin orbital period is assigned according to a fit to the 
rotational velocity data of Lang (1992). 
All aspects of standard binary evolution are treated according to the 
BSE algorithm (Hurley, Tout \& Pols 2002). 
Circularization of eccentric orbits and synchronization of stellar rotation 
with the orbital motion owing to tidal interaction is modelled in detail. 
Angular momentum loss mechanisms, such as gravitational radiation 
and magnetic braking, are also modelled. 
Wind accretion, where the secondary may accrete some of the material 
lost from the primary in a wind, is allowed with the necessary adjustments 
made to the orbital parameters in the event of any mass variations. 
Mass transfer also occurs if either star fills its Roche lobe and may proceed 
on a nuclear, thermal or dynamical timescale. 
In the latter regime the radius of the primary increases in response to 
mass loss at a faster rate than the Roche lobe of the star. 
Stars with deep surface convection zones and degenerate stars are 
unstable to such dynamical timescale mass loss unless the mass ratio of 
the system is less than some critical value. 
The outcome is a common-envelope (CE) event if the primary is a 
giant. 
This results in merging or formation of a close binary, or a 
direct merging if the primary is a WD or low-mass MS star. 
On the other hand, mass-transfer on a nuclear or thermal timescale 
is assumed to be a steady process. 
Prescriptions to determine the type and rate of mass transfer, the response of 
the secondary to accretion and the outcome of any merger events 
are in place in BSE and the details can be found in Hurley, Tout \& Pols (2002). 

One aspect that we should elaborate on here is the modelling of blue stragglers 
within BSE (and by association {\tt NBODY4}). 
If a MS star accepts mass from another star then it is rejuvenated. 
How this is done depends on whether the MS star has a radiative 
core ($0.35 \leq M/M_\odot \leq 1.25$), a convective core 
($M > 1.25 M_\odot$) or is fully convective ($M < 0.35 M_\odot$). 
As a MS star gains mass it evolves up along the MS to higher luminosity and effective 
temperature. 
If the core is radiative the fraction of hydrogen that has been burnt is only 
slightly affected so that the effective age of the star decreases. 
In practice the age of the star is altered so that the fraction of its MS 
lifetime that has elapsed is unchanged by the change of mass. 
For MS stars with convective cores, and fully convective stars, the core grows 
and mixes in unburnt fuel as the star gains mass, so that the star appears 
even younger. 
The rejuvenation process is approximated by conserving the amount of burnt 
hydrogen and assuming that the core mass grows directly proportional to the 
mass of the star. 
We then take the remaining fraction of MS lifetime to be directly proportional 
to the remaining fraction of unburnt hydrogen at the centre to set the new 
effective age of the star. 
Owing to the increase in mass the remaining MS lifetime of the star has been reduced 
but it has been rejuvenated relative to other stars of its new mass. 

The merger or collision of two MS stars produces a new MS star and we assume 
that the stellar material is completely mixed in the process, with no mass lost 
from the system. 
The age of the new star is calculated on the assumption that, 
as stars evolve across the MS 
core hydrogen burning proceeds uniformly and that the end of the MS is 
reached when 10 per cent of the total hydrogen has been burnt. 
These stars are also rejuvenated relative to other stars of the same mass. 
The process of rejuvenation and subsequent evolution of the MS star are most 
likely adequate in the case of steady mass transfer although there is a 
lack of detailed studies to confirm this (Sills et al. 2003). 
The assumption that no mass is lost in a collision is in fair 
accordance with smoothed-particle hydrodynamics 
simulations of low-velocity collisions (Sills et al. 2001), but 
only a limited amount of mixing is found in these simulations 
so our assumption of complete mixing is questionable, 
and even more so in the case of slow mergers.
In addition, the collision product is not initially in thermal equilibrium
and requires a thermal timescale to contract to its MS state. 
During this time it has a higher 
probability to undergo additional interaction (Lombardi et al. 2003). 
Furthermore the collision or merged product is, in most cases, rapidly rotating
and needs to lose angular momentum in order to contract 
(Lombardi, Rasio \& Shapiro 1996). 
These effects are not considered in our simplified model of post-collision 
BSs but would have an effect on the appearance of these stars. 

The great advantage of having the identical stellar and binary evolution algorithms in 
an $N$-body code and a population synthesis code is that we can evolve 
the same populations inside and outside the cluster environment to 
quantify how the dynamical evolution affects the stellar populations. 
$N$-body simulations presented in this work were conducted on the 
32-chip GRAPE-6 boards (Makino 2002) located at the 
American Museum of Natural History.

\section{Results for Model 1}
\label{s:result1}

\begin{figure}
\includegraphics[width=84mm]{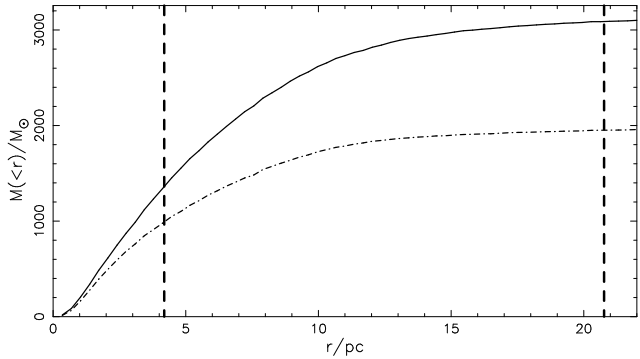}
\caption{
The mass profile of Model~1 at $4\,$Gyr. 
The solid line represents the total cluster mass and the 
dashed-dot line is the luminous mass. 
Dashed vertical lines show the tidal radius ($20.8\,$pc) and 
the half-mass radius for the luminous mass ($4.2\,$pc). 
\label{f:fig2}}
\end{figure}

The cluster of $9\,000$ single stars and $9\,000$ binaries set up in the manner 
described in Sections~\ref{s:initial} and \ref{s:binary} was evolved from 
zero-age to an age of $5.8\,$Gyr when $1\,000$ stars remained. 
This simulation took approximately one month to complete. 
Figure~\ref{f:fig2} shows the mass profile of the simulated cluster after 
$4\,$Gyr of evolution. 
The mass remaining in the cluster at this time is $3\,175 M_\odot$.  
It is contained within a tidal radius of $21\,$pc. 
We note that the mass profile does continue beyond the tidal radius because 
stars are not removed from the simulation until their distance from the density 
centre exceeds twice that of the tidal radius 
(cf. Terlevich 1987; Giersz \& Heggie 1994). 
For this particular model the mass exterior to the tidal radius is $84 M_\odot$, 
or 2.7\% of the total mass. 
Also shown in Figure~\ref{f:fig2} is the mass profile for the luminous 
mass which amounts to $1\,987 M_\odot$. 
The luminous mass, that in stars above $0.5 M_\odot$ burning nuclear fuel, 
contained within $10\,$pc is $1\,730 M_\odot$ and it 
has a half-mass radius of $3.0\,$pc. 
The half-mass radius for all stars is $4.9\,$pc. 
Fan et al. (1996) determined a luminous mass for M67 
of $1\,016 M_\odot$, rising to $1\,270 M_\odot$ when corrected for 
binaries, for stars within $10\,$pc of the cluster centre. 
Our model has too much mass remaining to be a good representation 
of M67. 
It is not until $5.2\,$Gyr that we find $M_{\rm L} = 1\,000 M_\odot$ and we 
consider this too old to be relevant to M67. 
General results for this model are summarized in Table~\ref{t:table2}. 

At $4\,$Gyr the model contains only one blue straggler. 
This star resides in a non-primordial binary with an orbital period of 
$83\,$d and an eccentricity of 0.5. 
The mass of the BS is  $1.5 M_\odot$ -- compared to 
$M_{\rm TO} = 1.32 M_\odot$ -- and the companion is a 
$0.64 M_\odot$ MS star. 
Slightly earlier ($3\,950\,$Myr) there were two BSs and slightly later 
($4\,050\,$Myr) there are three, so an average of two BSs at this age. 
The most BSs observed in the cluster at any one time, 
from an age of $3\,$Gyr onwards,  
is $5$ at $4.5\,$Gyr. 
So the combination of the Kroupa (1995b) setup for the 
primordial binaries and the dynamical evolution of the cluster does not 
explain the actual M67 BS population. 
This is not a quirk of the model, either in terms of the initial conditions 
or statistics. 
We have performed a variety of simulations with the primordial binaries 
chosen in the same manner but have altered various 
characteristics of the starting model such as the King model for the density profile 
and the tidal radius filling factor. 
These models were evolved as part of a broader project to understand the 
evolution of binary-rich open clusters and will be described in detail in 
a forthcoming paper.
In none of these models did we see substantial BS production. 
It has been noted previously (Giersz \& Heggie 1994) that statistical fluctuations 
in $N$-body models may be reduced by averaging over the results of 
many simulations. 
These fluctuations are amplified for small $N$ and should not be of major 
concern in models of the size presented here (compared to the $N = 500$ 
models discussed by Giersz \& Heggie 1994). 
However, we have repeated Model~1 with an identical set up except for 
a different seed for the random number generator. 
The evolution is similar -- the mass remaining in the cluster after $4\,$Gyr 
is $2\,957 M_\odot$ within a tidal radius of $20\,$pc and the luminous mass 
within the central $10\,$pc is $1\,692 M_\odot$. 
At no time subsequent to an age of $3\,$Gyr are more than five BSs found 
in the model cluster. 
Therefore the nature of the primordial binaries in this type of simulation does not 
seem appropriate for a model of M67.  
Rather than continuing further with our analysis of Model~1 we 
instead turn our attention to Model~2.

\section{Results for Model 2 -- the M67 model}
\label{s:result2}

A flaw of Model~1 was that it contained too much mass at $4\,$Gyr to 
be considered a good model of M67. 
One way to reduce the mass remaining in a cluster at a certain age is 
to increase the strength of the tidal field in which the cluster evolves 
(Vesperini 1997; Baumgardt 2001). 
This reduces the tidal radius of the cluster which leads to an increase 
in the escape rate of stars. 
Associated with the reduction in mass is a lowering of the relaxation 
timescale (typically the half-mass radius will also be smaller) so that 
the cluster is dynamically older for a given physical age. 
The parameters given for the orbit of M67 in the Galaxy (Carraro \& Chiosi 1994) 
indicate an eccentricity of 0.14 and a time-averaged semi-major axis of 
$8\,$kpc. 
So, when evolving Model~2, we will place the cluster on an orbit at 
$8.0\,$kpc from the Galactic centre. 
An orbital speed of $220\, {\rm km} \, {\rm s}^{-1}$ is still applicable at this 
distance (Chernoff \& Weinberg 1990) so this remains unchanged and we 
keep the cluster on a circular orbit. 

With the stronger tidal field we now expect the mass remaining after $4\,$Gyr 
of evolution to be less than that found for Model~1 and closer to the mass 
of M67, as desired. 
However, we anticipate that the shorter relaxation time will drive the 
evolution faster than required to give the desired ~25\% reduction in 
cluster mass. 
So we make an associated increase in the particle number of the 
starting model to $12\,000$ single stars and $12\,000$ binaries. 
This gives $M_0 = 18\,700 M_\odot$ within a tidal radius of $32\,$pc. 

The only other change in Model~2 compared to Model~1 is in the 
set-up of the primordial binary population. 
As discussed in Section~\ref{s:binary} a flat distribution of $\log a$ will 
be used to generate the orbital separations of the primordial binaries.  
This has a cutoff at $50\,$au which is comparable to the hard-soft binary 
limit (Heggie 1975) of about $40\,$au for Model~2 with an initial half-mass 
radius of $3.9\,$pc. 
In all other respects the set up for Model~2 is the same as for Model~1. 
The two starting models are compared in Table~\ref{t:table1}. 
From the $12\,000$ primordial binaries in Model~2 we should 
expect 25 BSs from binary evolution alone (see Section~\ref{s:binary}). 

\begin{figure}
\includegraphics[width=84mm]{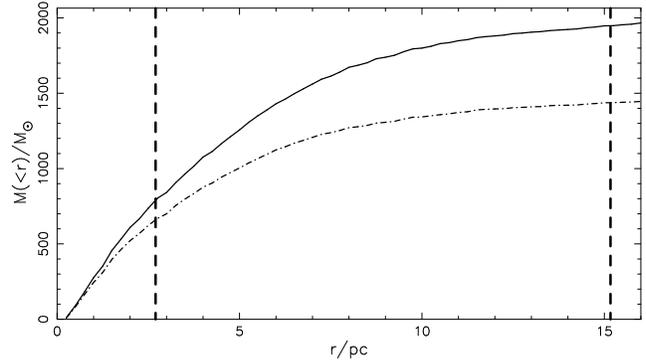}
\caption{
The mass profile of Model~2 at $4\,$Gyr. 
The solid line represents the total cluster mass and the 
dashed-dot line corresponds to the luminous mass. 
Dashed vertical lines show the tidal radius ($15.2\,$pc) and 
the half-mass radius for the luminous mass ($2.7\,$pc). 
\label{f:fig3}}
\end{figure}

The mass profile for Model~2 at $4\,$Gyr is shown in Figure~\ref{f:fig3}. 
It has a total mass of $2\,037 M_\odot$ and a luminous mass 
of $1\,488 M_\odot$ within a tidal radius of $15\,$pc. 
The luminous mass within $10\,$pc of the cluster centre ($M_{{\rm L},10}$) is now $1\,342 M_\odot$ 
which compares well with the binary corrected value found by Fan et al. (1996). 
Slightly later in the evolution ($4.1\,$Gyr) the model has 
$M_{{\rm L},10} = 1\,181 M_\odot$ with a tidal radius of $14.5\,$pc. 
The half-mass radius of MS stars observed within $10\,$pc was determined 
by Fan et al. (1996) to be $2.5\,$pc and our model has $2.7\,$pc for the same 
set of stars. 
So we now have a good match to the observed mass of M67 
as well as the half-mass and tidal radii. 
General results for this model are summarized in Table~\ref{t:table2}. 

\begin{figure}
\includegraphics[width=84mm]{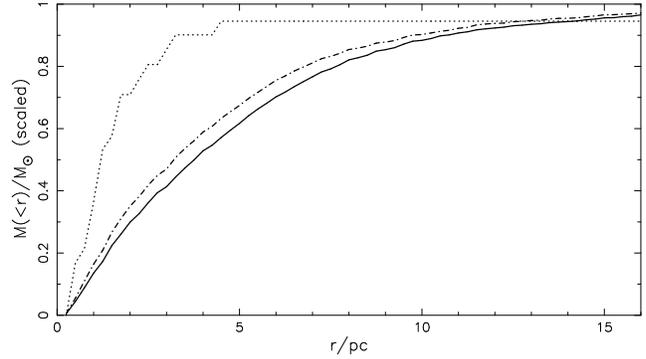}
\caption{
Mass profiles for Model~2 at $4\,$Gyr scaled by the total 
mass involved in constructing the profile,  
all stars (solid line), luminous mass 
(dashed-dot line) and the blue straggler mass (dotted line). 
\label{f:fig4}}
\end{figure}

In Figure~\ref{f:fig4} we once again show the mass profiles at $4\,$Gyr 
but this time normalized to the total mass of each profile. 
The luminous mass is more centrally concentrated. 
Naively this result would be expected owing to mass-segregation when we 
consider that low-mass MS stars are excluded from the luminous mass. 
However, the effect is counteracted somewhat by the added exclusion of white 
dwarfs (WDs) from the luminous mass because these are generally born in the central 
regions of the cluster and themselves are centrally concentrated relative to the 
entire population. 
Also shown in Figure~\ref{f:fig4} is the normalized mass profile of the BSs of 
which there are 20 in the model at $4\,$Gyr. 
These are more likely to be found in the centre of the cluster and 
have a half-mass radius of $1.1\,$pc. 
Fan et al. (1996) determined a half-mass radius of $1.6\,$pc for the M67 BSs. 
We elaborate further on the BSs and other stellar populations in 
Section~\ref{s:res_sub2}. 

\begin{figure*}
\includegraphics[width=168mm]{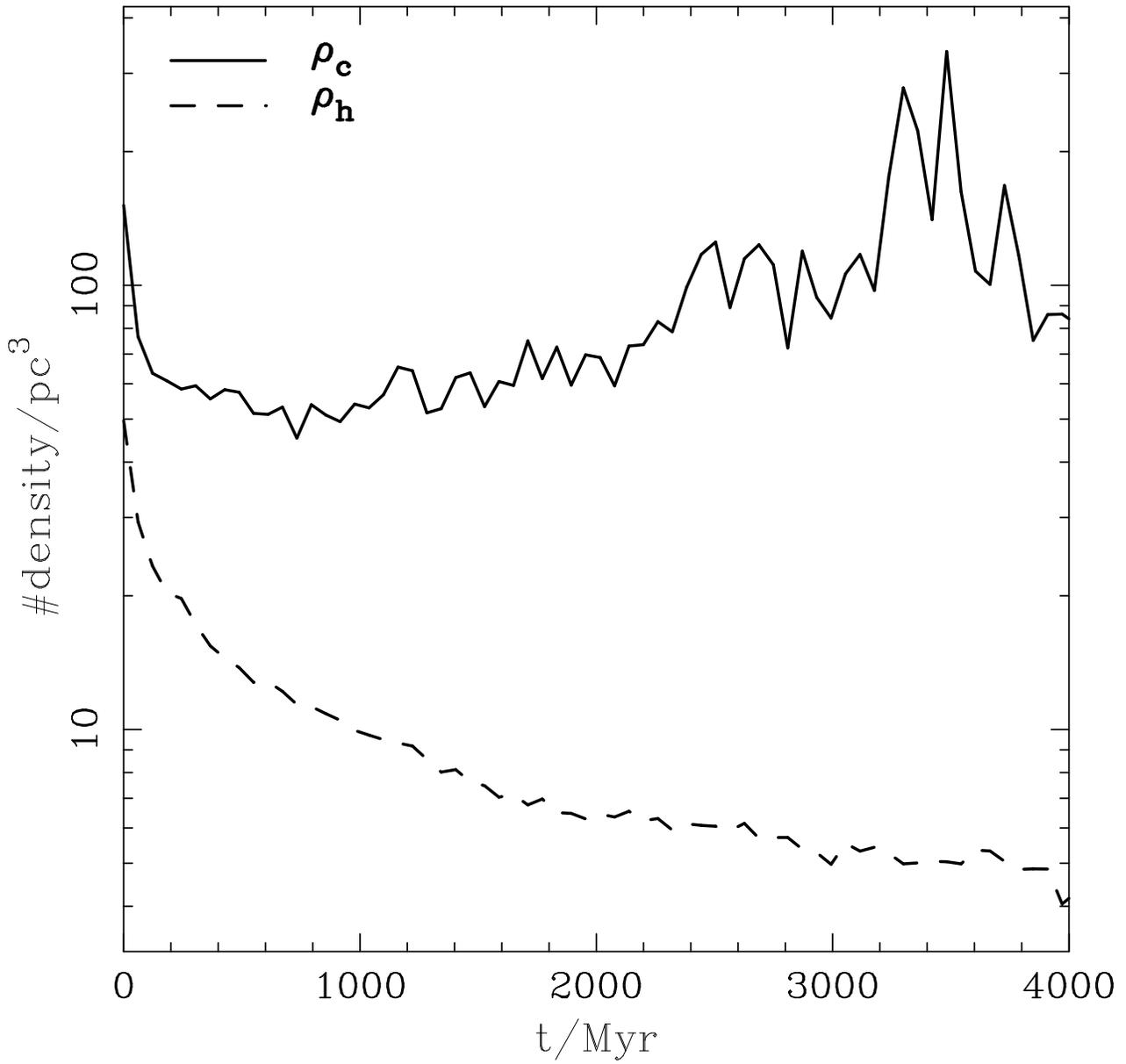}
\caption{
Evolution of the number density of stars within the core (solid line) 
and the half-mass radius (dashed line) for Model~2. 
\label{f:fig5}}
\end{figure*}

The cluster at $4\,$Gyr of age is a dynamically relaxed system 
-- 13 half-mass relaxation times have elapsed in reaching this point. 
Figure~\ref{f:fig5} shows the behaviour of the number density of 
stars contained within the core and the half-mass radius of the cluster 
as it evolved from $0$ to $4\,$Gyr. 
The core density of the starting model was $150\, {\rm stars} \, {\rm pc}^{-3}$. 
This rose to a maximum of $330\, {\rm stars} \, {\rm pc}^{-3}$ after 
$3.5\,$Gyr. 
Here the core radius is the density-weighted value commonly used in 
theoretical models (Casertano \& Hut 1985; Aarseth 2003) which is 
typically smaller than the core radius determined from observational 
techniques (Wilkinson et al. 2003). 
For binary-rich clusters there is no clear core collapse, at least not 
to the extent that we witness in simulations without primordial binaries,  
where a high core density is required for binary formation (e.g. Hurley et al. 2004). 
Heating of the core by binaries occurs from the beginning in simulations 
with a large primordial binary population and this helps to keep the 
evolution of the core radius relatively regular and to avoid extreme 
fluctuations in central density. 
Stellar evolution mass-loss also leads to core expansion, 
especially at early times when massive stars evolve away from the MS. 
The core density of the model at $4\,$Gyr is $83\, {\rm stars} \, {\rm pc}^{-3}$. 
The density of stars contained within the half-mass radius started at 
$50\, {\rm stars} \, {\rm pc}^{-3}$ and evolved to $4\, {\rm stars} \, {\rm pc}^{-3}$ 
at $4\,$Gyr. 
Hence this was not a particularly dense system. 
At an age of $5\,$Gyr the model cluster contains only 200 stars with a total mass
of $260 M_\odot$ and has almost reached the point of complete dissolution. 
In this paper the focus is on the results of the simulation at 
the age of M67 and the relevance of the model in improving our 
understanding of the observed properties of M67. 
We do not dwell on the details of the simulation in reaching an 
age of $4\,$Gyr. 
We shall present the long-term evolution of a binary-rich star cluster in 
another paper (Hurley et al., in preparation).

\subsection{Cluster Structure}
\label{s:res_sub1}

\begin{figure*}
\includegraphics[width=168mm]{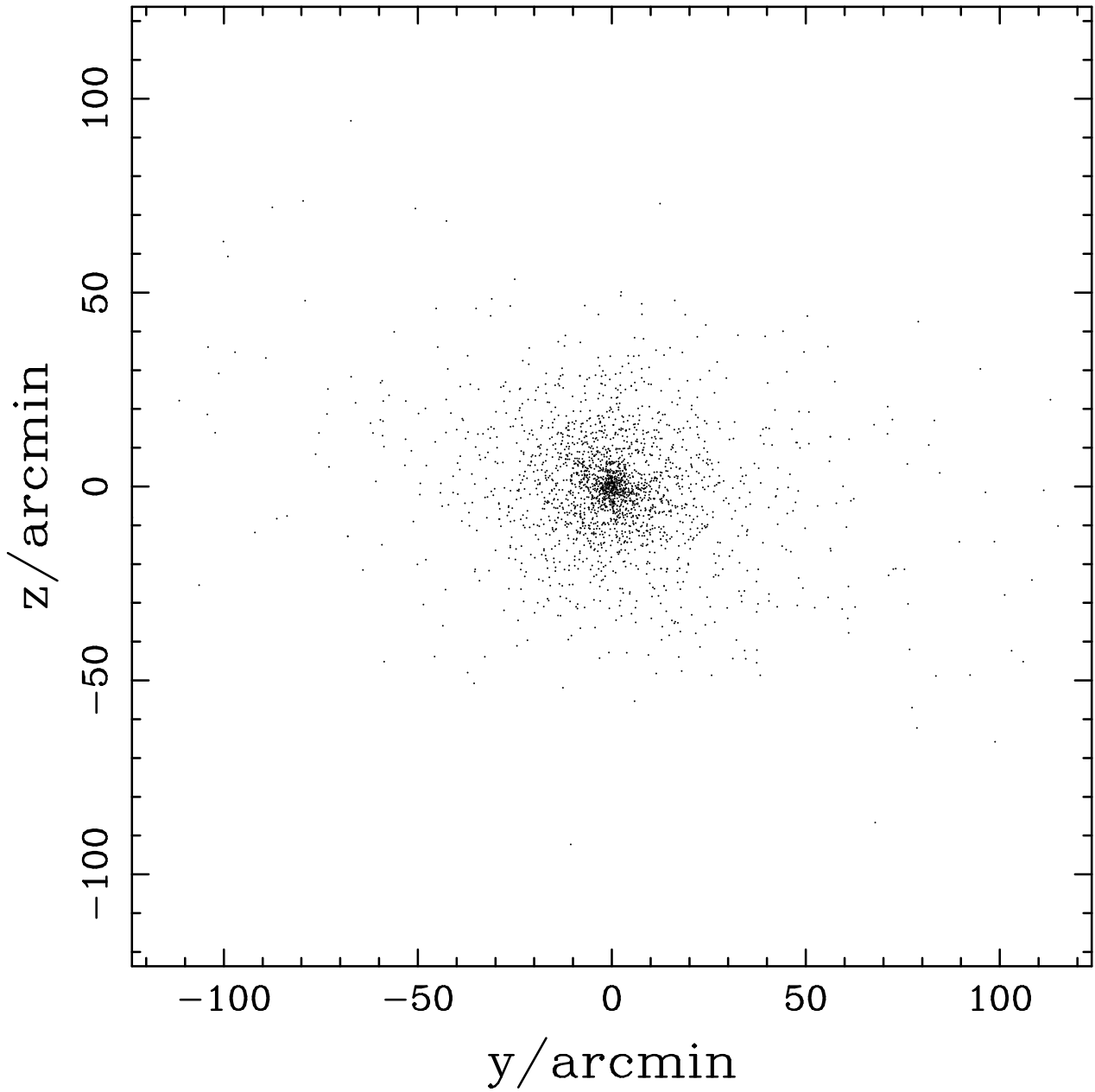}
\caption{
Model~2 at $4\,$Gyr shown in the observed plane 
(the yz- or transformed YZ-plane).  
\label{f:fig6}}
\end{figure*}

The coordinate system used in the $N$-body simulation has the cluster centre-of-mass 
as the origin, the X-axis directed away from the Galactic centre, the Y-axis in the 
direction of rotation about the Galactic centre and the Z-axis normal to the plane of the 
disk. 
In order to show how the model of M67 might appear on the sky we have rotated the 
model according to the Galactic coordinates of M67 ($l = 215.68^\circ$, $b = 31.93^\circ$: 
Bonatto \& Bica 2003) so that the transformed YZ-plane becomes the observed plane 
(which we have also denoted the yz-plane). 
This view of the model is presented in Figure~\ref{f:fig6}. . 

\begin{figure}
\includegraphics[width=84mm]{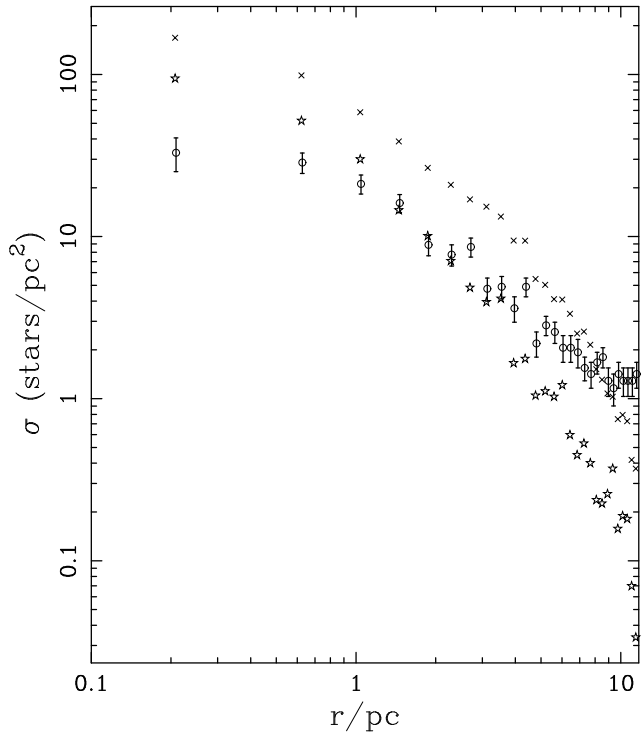}
\caption{
Surface density profile of M67 from 2MASS data 
provided by Bonatto (private communication: open circles). 
Profiles from the model for all stars ($\times$ symbols) and for only luminous 
stars with mass greater than $0.8 M_\odot$ (open star symbols) 
are also shown. 
The distributions have not been normalized. 
$1 \sigma$ error bars have been included for the observed profile but for 
the sake of clarity we have not included error bars for the model points 
although the errors will be of comparable magnitude. 
\label{f:fig7}}
\end{figure}

Bonatto \& Bica (2005) used 2MASS data to construct a surface density profile 
for M67. 
This is shown in Figure~\ref{f:fig7}. 
The limiting radius for the observations is $11.7 \pm 0.6\,$pc and 
fitting a King (1966) surface density profile to the data gives a core 
radius of $1.14 \pm 0.13\,$pc within a tidal radius of $16 \pm 3\,$pc 
(Bonatto \& Bica 2005). 
Using the same radial bins as for the observed M67 surface density profile,  
we also show the profile for the model in Figure~\ref{f:fig7}. 
The surface density of stars is greater in the model, especially in the central 
regions. 
However, the observed profile includes only stars with $J < 14.5$ which 
corresponds to stars with masses of about $0.8 M_\odot$ or greater. 
Restricting the model profile to stars in this range, we find that agreement 
between the model and the observations is better. 
The match is good in the $1-6\,$pc range which is the region that contains the 
half-mass radius of the cluster. 
Exterior to $6\,$pc the observed surface density drops to levels similar to that 
of the background counts (about $1\, {\rm star} \, {\rm pc}^{-2}$) so it is not 
significant that the model surface density is lower in this region. 
We note that background contamination is nil in the $N$-body model. 
At the centre the model surface density is still too high. 
The model is overdense by a factor of 3 in the central bin. 
Fitting a King (1966) surface density profile to the model data gives a 
core radius of $0.6\,$pc so the model is clearly more centrally concentrated 
than observations indicate for M67. 
Saturation effects could be lowering the observed star counts in the core of 
M67 but this is not expected to be significant. 
We note that there is no noticeable change in the model profile if we use 
different projections to construct it. 

\begin{figure}
\includegraphics[width=84mm]{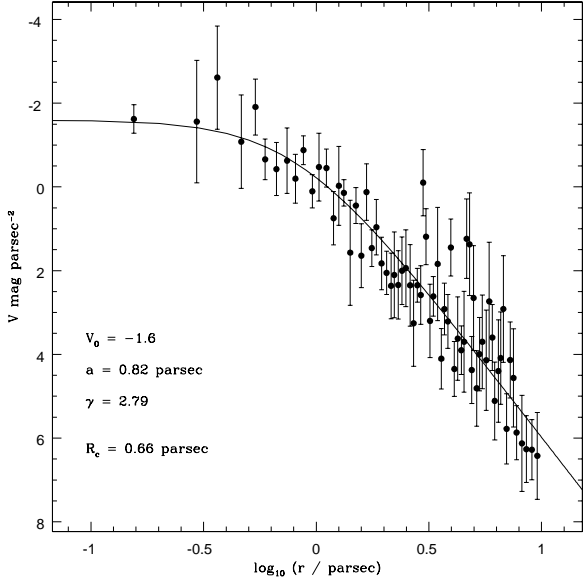}
\caption{
Surface brightness profile for Model~2 at $4\,$Gyr (solid data points) 
constructed using software provided by Mackey \& Gilmore (2003). 
The projection is along the Y-axis. 
Error bars are calculated according to the method outlined
by Djorgovski (1988, in IAU Symp. 126, p333) where a given annulus is split
into eight sectors of equal area. 
The internal error for the annulus is
the standard deviation of the surface brightness values for the eight
sectors.
A fit to the data of an Elson, Fall \& Freeman (1987) model 
(solid line) gives a core radius of $0.66\,$pc. 
This is a three-parameter model $V(r) = V_0 (1 + r^2/a^2)^{- \gamma/2}$ 
where $V_0$ is the central surface brightness and $a$ is related to the 
core radius by $r_{\rm c} = a (2^{2/\gamma} - 1)^{1/2}$. 
The $\chi^2$ error of the fit is 1.05 which represents the sum of the squares of the 
differences between the data and model points in each bin weighted by the 
error for that bin. 
\label{f:fig8}}
\end{figure}

\begin{figure}
\includegraphics[width=84mm]{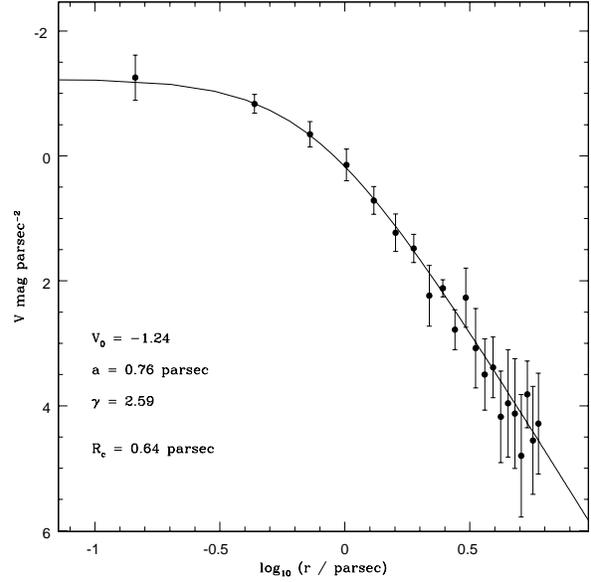}
\caption{
Same as Figure~\ref{f:fig8} but restricted to stars in the range 
$12 < V < 17$. 
The main-sequence turn-off is at $V \simeq 13$ (corresponding to 
a mass of $1.32 M_\odot$) so the range covers one magnitude 
above the turn-off, approximately half-way up the giant branch, 
and four magnitudes below the turn-off, 
down to a mass of $0.75 M_\odot$. 
The surface density profile fit has a scaled $\chi^2$ error of 0.25 
and gives a core radius of $0.64\,$pc. 
\label{f:fig9}}
\end{figure}

A common method used in the analysis of observed cluster data is to construct 
a surface brightness profile. 
This can also be done with the $N$-body cluster data for which the mass, 
luminosity, stellar radius and position of each star is known, as well as 
whether or not the star is in a binary. 
First we calculate magnitudes and colours for each star using bolometric 
corrections provided by Kurucz (1992) and, in the case of WDs, 
Bergeron, Wesemael \& Beauchamp (1995). 
Then the magnitude and position information is passed through the pipeline 
described by Mackey \& Gilmore (2003) in their analysis of LMC clusters 
to give a projected surface density profile. 
The resulting data points for our M67 model for the V-band in the XZ-plane are 
shown in Figure~\ref{f:fig8}. 
A fit of the three-parameter Elson, Fall \& Freeman (1987) model to the cluster 
profile is also shown and this gives a core radius of $0.66\,$pc. 
In this case all stars have been considered and the data is rather noisy. 
If we repeat the process but remove all stars with $V < 12$ in order to reduce 
saturation effects by bright stars and also remove all stars with $V > 17$ 
to mimic a faint detection limit then we get the much cleaner profile shown 
in Figure~\ref{f:fig9}. 
The fit to these data points gives a core radius of $0.64\,$pc.

\subsection{Stellar populations}
\label{s:res_sub2}

In Figure~\ref{f:fig10} we present the colour-magnitude diagram (CMD) 
of Model~2 at $4.0\,$Gyr. 
There are $870$ single stars and $1\,325$ binaries in the cluster at 
this time. 
The MS turn-off mass is $1.32 M_\odot$. 
The cluster contains $2\,968$ MS stars, 57 giants and subgiants, 
2 naked helium (nHe) stars, 491 WDs and 2 neutron stars. 
Defining BSs as MS stars with mass in excess of the MS turn-off mass 
(by 2\% or more) 
we find 20 with masses in the range $1.4 - 2.1 M_\odot$. 
These form the group of stars blueward of the the MS in the CMD. 
Nine of the BSs are in binaries. 
The BS population is discussed in detail in Sect.~7.2.1.
There are six RS Canum Venaticorum 
(RS CVn) stars in the cluster. 
These are binaries with periods of $1 \le P / {\rm d} \le 14$ that contain a 
cool subgiant star and a MS companion (Hall 1976).  
They are believed to be 
sources of X-ray emission. These and other expected X-ray sources
in the cluster are discussed in Sect. 7.2.2.

In the cluster at $4\,$Gyr are 226 single WDs. 
Most of these lie on the distinctive WD cooling track seen in the lower 
left corner of the CMD. 
There are 60 double-WD binaries, the majority of which 
are found in the CMD by their position above the sequence of 
standard single WDs. 
Of the double WDs, 28 have periods less than $1\,$d. 
The stars appearing in the region between the WD sequence and the MS 
are WD-MS star binaries in which the WD is still young and fairly hot. 
As the WD cools the binary moves across the CMD towards the MS as 
the MS star begins to dominate the appearance of the binary. Further
particulars of the WD population are discussed in Sect. 7.2.3.
There is also one cataclysmic variable (CV) in the cluster at this time, 
located at $V = 23.06$ and $(B-V) = 0.53$ in the CMD. 
It comprises a $0.1 M_\odot$ MS donor star in a $0.55\,$d circular orbit 
with a $0.3 M_\odot$ helium WD. 
The CV phase began at an age of $1.9\,$Gyr when the MS star mass 
was $0.18 M_\odot$. 

A summary of selected stellar population results for the M67 model is 
given in Table~\ref{t:table3} along with results from Model~1. 
There are two extremely blue stars at $V \approx 12$ in the CMD. 
These are not BSs but binaries comprised of evolved nHe 
stars with MS star companions. 
Both evolved from primordial binaries in which the nHe star 
was produced in a CE event initiated by the
progenitor of the nHe star filling its Roche lobe while on the early AGB. 
However, the fainter of the two binaries would not exist in this state 
without having experienced a significant perturbation to its orbit. 
It originated as a primordial binary with an eccentricity of $0.5$ and a 
period of $30\,902\,$d. 
Evolved in isolation the two stars would not have become close enough 
to interact and the result would be a WD-MS star binary with an 
orbital period of about $300\,$yr at $4\,$Gyr. 
Within the cluster environment the binary received a perturbation 
to its orbit from a third star, in a flyby encounter after $350\,$Myr, 
while still a MS-MS star binary. 
This resulted in a slight decrease in orbital period but more important 
pumped the eccentricity up to 0.95 and subsequent tidal evolution 
brought the stars close enough for mass-transfer to begin. 
The CE event then reduced the orbital period further 
so that the nHe-MS star binary observed at $4\,$Gyr has $P = 12.6\,$d 
and is circular. 
The other nHe-MS star binary has a period of $3.5\,$d. 

Another anomalous star in the CMD lies towards the base of the 
giant branch (GB) but below the subgiant branch at $V = 13.32$ 
and $(B-V) = 0.83$. 
This is a single star with a mass of $1.91 M_\odot$ recently created 
in a CE merger event. 
At an age of $3.8\,$Gyr a $1.37 M_\odot$ subgiant primary filled its 
Roche lobe and began steady Case~B mass transfer on to its 
$0.83 M_\odot$ MS star companion. 
When the primary reached the end of the subgiant phase 
(about $50\,$Myr later) its mass had been reduced to $0.98 M_\odot$ 
and the companion mass was $1.22 M_\odot$ owing to the mass-transfer 
being conservative. 
At this stage the envelope of the primary was fully convective and the 
entire envelope overflowed the Roche lobe on a dynamical timescale 
to create a common envelope. The mass ratio inversion had 
not been enough to avoid this.  
The giant core and the MS star spiralled together and expelled $0.29 M_\odot$ 
of the envelope via dynamical friction before merging to form a $1.91 M_\odot$ 
giant with a core mass of $0.16 M_\odot$. 
This star is more massive than the normal giants in the cluster at this time 
but has a core mass less than expected for a giant of this mass and in 
fact less than the core mass of the stars residing at the base of the cluster GB. 
The reduced core mass is the result of the mass loss experienced by the 
subgiant progenitor which restricted its core mass growth as it evolved 
across the subgiant branch. 
This is the cause of the merged giant lying below the subgiant branch. 
The increased mass of the giant will cause it to remain on the blue side 
of the standard cluster GB as it continues its giant evolution.

\subsubsection{Blue stragglers}

Table~\ref{t:table4} provides a list of the 20 BSs in the M67 model. 
All of the eleven single BSs were produced from the merger of 
two MS stars in a primordial binary. 
In nine of these cases BS formation was via 
the onset of Case~A mass-transfer and the eventual coalescence 
of the stars as angular momentum was removed from the system.  
For seven of these the evolution proceeded as if the 
binaries were evolved in isolation. 
The cluster environment did not interfere. 
In one case (star \#2411), a perturbation to the orbit hastened the onset of
Case~A mass-transfer so that a BS that would not have been created until 
after $4\,$Gyr was formed earlier and observed at $4\,$Gyr. 
In another case (\#3021) mass transfer was delayed by the involvement of 
the binary in a temporary exchange interaction which increased the 
orbital period. 
This BS would not have been observed at $4\,$Gyr without the interference of 
the third star as it would have already evolved to become a WD. 
On the other hand, if the exchange interaction had caused a larger period increase 
the BS may have formed after $4\,$Gyr or not at all. 
The remaining two of the single BSs formed from initially long-period 
binaries that had their eccentricity pumped up to 0.99 by weak flyby encounters. 
This led to a collision of the MS star components at periastron and creation 
of a BS. 
In neither binary would the component stars have become close enough 
to interact and merge without the intervention of a third star. 
Descriptions of the evolution pathways for the four single BSs that were affected 
in some way by the cluster environment are given in Table~\ref{t:table5}. 

Descriptions for the nine binary BSs are also given in Table~\ref{t:table5}. 
Two of these are primordial binaries but both have been affected by 
interactions with other cluster members. 
This explains the eccentric nature of the orbits. 
In one case the BS was formed in a wide circular binary as a result of 
Case~C mass-transfer and a subsequent perturbation induced the 
eccentricity observed in the orbit. 
The other primordial binary became involved in a four-body interaction 
and one of the binary components collided with another member of the subsystem 
to form the BS which then emerged bound to its original companion. 
The remaining BS-binaries are non-primordial and their formation involved 
an exchange interaction at some point. 
In four of these cases a primordial binary became part of a three- or four-body 
system and perturbations to the orbit drove the eccentricity of the binary 
towards unity so that the stars collided. 
The merged star, a BS, was then exchanged into a binary. 
One of these cases then underwent a second collision in an eccentric 
binary while another was subsequently perturbed. 
A BS binary was formed from Case-C mass transfer in a primordial binary 
but the BS was then exchanged into a new binary and, just prior to $4\,$Gyr, 
Case~B mass-transfer began in this binary. 
This further increased the mass of the BS. 
Another case saw an exchange interaction form a binary which evolved 
to a state of Case-A mass transfer followed by coalescence to a 
BS which was later exchanged into a new binary. 
The final case involved an exchange interaction followed by two 
collision events. 
So the binary BSs were formed by a variety of means and this 
resulted in a mix of orbital parameters: 
short-period and circular, short-period (less than $1\,000\,$d) and eccentric, 
long-period (greater than $1\,000\,$d) and eccentric (see Table~\ref{t:table4} 
for full details). 

Only seven of the 20 BSs in the model evolved from unperturbed primordial
binaries. Of the remaining 13, eight BSs were formed by collisions that were
induced by three- or four-body interactions, or by perturbations that drove
up the eccentricity to almost unity. These eight could not have formed outside
the cluster environment and the same is true for the case~A merged star that formed
after an exchange. The other four would have become BSs by binary
interaction alone but two of these would not have been observed as BSs 
at an age of $4\,$Gyr. Hence, for approximately half the BS population the dynamical
cluster environment was instrumental in producing them.

As mentioned earlier the half-mass radius for the BSs in the M67 model 
is $1.1\,$pc. 
This is much less than the half-mass radius for all stars 
($3.8\,$pc) and 
indicates that the BSs are a centrally concentrated population 
which is in agreement with their observed distribution in M67 (Fan et al. 1996). 
The fact that BSs are more likely to be found in the centre of the cluster 
than the outer regions is confirmed by Figure~\ref{f:fig11} which splits 
the CMD into two regions, stars contained within the half-mass radius 
and stars exterior to this. 
We see that only two of the BSs in the model at $4\,$Gyr are found outside 
of the half-mass radius and one of these was situated well within the 
half-mass radius when it formed\footnote{This BS lies outside of the tidal 
radius at $4\,$Gyr, $19.4\,$pc from the cluster centre. It is counted as a 
cluster member because its orbit is such that it remains bound to the 
cluster and subsequently returns within the tidal radius. 
The particulars of this BS are provided in Table~\ref{t:table4} (BS \#2565) and its 
evolution pathway is detailed in Table~\ref{t:table5}. 
The radial position of this BS explains why the normalized mass profile of the 
BSs in Figure~\ref{f:fig4} does not reach unity within the tidal radius.}. 
The reasons for this are twofold. 
Primordial binaries which are to produce a BS must have a mass 
in excess of the current MS turn-off mass ($1.3 M_\odot$) which 
itself is greater than the average mass of the cluster stars 
($0.6 M_\odot$). 
So the process of mass-segregation acts to concentrate these binaries at 
the centre of the cluster and the BSs themselves, when created, also 
tend to sink towards the centre. 
We also have BSs created in binaries formed from exchange interactions 
and these interactions are more likely in regions of high density 
(Heggie, Hut \& McMillan 1996) which favours the centre of the cluster 
(see Figure~\ref{f:fig5}). 

The 20 BSs in the model at $4\,$Gyr is less than the 25 predicted when 
evolving the primordial binaries in isolation (see Section~\ref{s:binary}). 
Considering that only 11 of the 20 BSs would have formed without help from
the cluster environment, this means that 14 potential BSs at $4\,$Gyr were lost 
during the simulation. 
So in addition to creating BSs the cluster environment is just as 
productive in destroying potential BSs, maybe even more so. 
The majority of {\it lost} BSs were the result of perturbations to primordial 
binaries that hardened the orbit and brought forward the onset of 
Case-A mass transfer. 
BS formation still occurred in these binaries but much earlier than it 
would have otherwise and this caused the BS phase to have ended 
prior to $4\,$Gyr. 
This is compensated to some extent by somewhat wider binaries being
hardened into orbits that allow them to experience Case-A mass transfer
and coalesce but otherwise would not have done so within 4 Gyr, and
by interactions that delay the onset of mass transfer as in the case of \#3021.
Other possibilities include an exchange interaction destroying the 
primordial binary or the binary being ejected from the cluster. 
Both of these events are rare compared to the hardening scenario.  
The former because the pre-exchange binary is short-period and the 
latter because the binary is relatively heavy. 

There are no BSs in long-period (nearly) circular binaries present at $4\,$Gyr 
whereas there are two observed in M67: $P = 1\,221\,$d, $e = 0.09$ 
and $P = 1\,154\,$d, $e = 0.07$ (Latham \& Milone 1996). 
However, BSs in such binaries were present in the simulation at other times. 
An example is a $1.8 M_\odot$ BS in a circular orbit of period $1\,445\,$d 
with a WD companion at $2.6\,$Gyr. 
The proto-BS initially accreted mass from the stellar wind of its AGB-star 
companion and then grew even more in mass with the onset of Case-C 
mass transfer, with dynamical timescale mass-transfer and CE  
evolution avoided because the stellar wind had significantly decreased the 
mass of the AGB star by this stage.  
Mass transfer ceased when the AGB star exhausted its envelope and became 
a WD. 
Also, BS \#1378 observed at $4\,$Gyr in a long-period eccentric binary was 
originally in a wide circular binary when formed at $T = 2\,$Gyr. 
It would have been found in this state at $4\,$Gyr if the orbit had not been 
perturbed by a passing star at $2.5\,$Gyr. 
So in a sense it is simply bad luck that we did not observe any BSs in long-period 
circular binaries after $4\,$Gyr of evolution. 
The same can be said for the incidence of super-BSs. 
M67 is observed to have a super-BS of mass $3 M_\odot$ (Leonard 1996) 
but our model at $4\,$Gyr does not contain any BSs 
with mass greater than twice the MS turn-off mass of the cluster. 
Super-BSs do form in the simulated cluster, 
a total of five for the entire simulation including one at $3.9\,$Gyr when 
the model contained 22 BSs. 
This super-BS of mass $3.2 M_\odot$ was found in a binary of period $4.6\,$d 
and eccentricity 0.45 with a $1.2 M_\odot$ MS star companion. 
The star began its life as a $1.2 M_\odot$ MS star with a $1.3 M_\odot$ 
companion in a circular primordial binary with $P = 1.2\,$d. 
After $2.7\,$Gyr the more massive MS star filled its Roche lobe and Case-A 
mass transfer proceeded until $3.1\,$Gyr when the stars coalesced to form 
a $2.5 M_\odot$ single BS. 
The BS was situated well within the core of the cluster ($0.1\,$pc from the centre) 
and at $3.4\,$Gyr exchanged itself into an existing binary. 
Its new companion was a $0.7 M_\odot$ MS star and the orbit had an 
eccentricity of 0.1 and a period of $1.3\,$d. 
The orbit quickly circularized owing to tidal forces and Case-A mass transfer 
began at $3.6\,$Gyr with coalescence almost immediately afterwards. 
This newly formed $3.2 M_\odot$ super-BS was exchanged into a wide 
binary with a $1.3 M_\odot$ companion at $3.7\,$Gyr and this binary was 
then involved in a short-lived four-body system that created the short-period 
eccentric binary observed at $3.9\,$Gyr. 
The super-BS evolved off the MS about $50\,$Myr later and just missed being 
included in the M67 model. 
This evolution example also serves to highlight how BSs may be formed in 
short-period eccentric binaries. 
M67 is observed to contain a BS in an orbit with period $4.18\,$d and 
eccentricity of 0.2 (Milone \& Latham 1992). 

The orbital parameters of all BS--binaries created in Model~2 at an age of 
$2\,$Gyr or later are shown in Figure~\ref{f:fig12}. 
Also shown are the six M67 BS-binaries with known orbital solutions. 
Figure~\ref{f:fig13} shows the total number of BSs and BSs within binaries 
as a function of cluster age up to $4.2\,$Gyr. 
We note that the numbers we have focussed on at $4\,$Gyr are typical of 
the cluster over the preceeding Gyr or so. 
The increase in BS number seen after $2\,$Gyr corresponds to an increase 
in central density as the cluster becomes dynamically more evolved. 
The BSs found in binaries during this timeframe are more likely to be in 
non-primordial binaries which is the opposite of what is found earlier in the 
evolution. 
In Figure~\ref{f:fig13} we also show the number of BSs in circular binaries 
with $P < 100\,$d. 
Only one BS binary of this type is found in the cluster at $4\,$Gyr but at 
earlier ages these binaries, formed primarily from Case B mass transfer, 
were dominant. 
Of the BS binaries formed prior to an age of $2\,$Gyr 82 per cent were circular 
with an orbital period of $100\,$d or less which contrasts with 23 per cent 
of the BS binaries in Figure~\ref{f:fig12} being of this type. 
The number of these binaries declines with age and this is 
linked to the destruction of primordial binaries as the cluster ages. 

\begin{figure*}
\includegraphics[width=168mm]{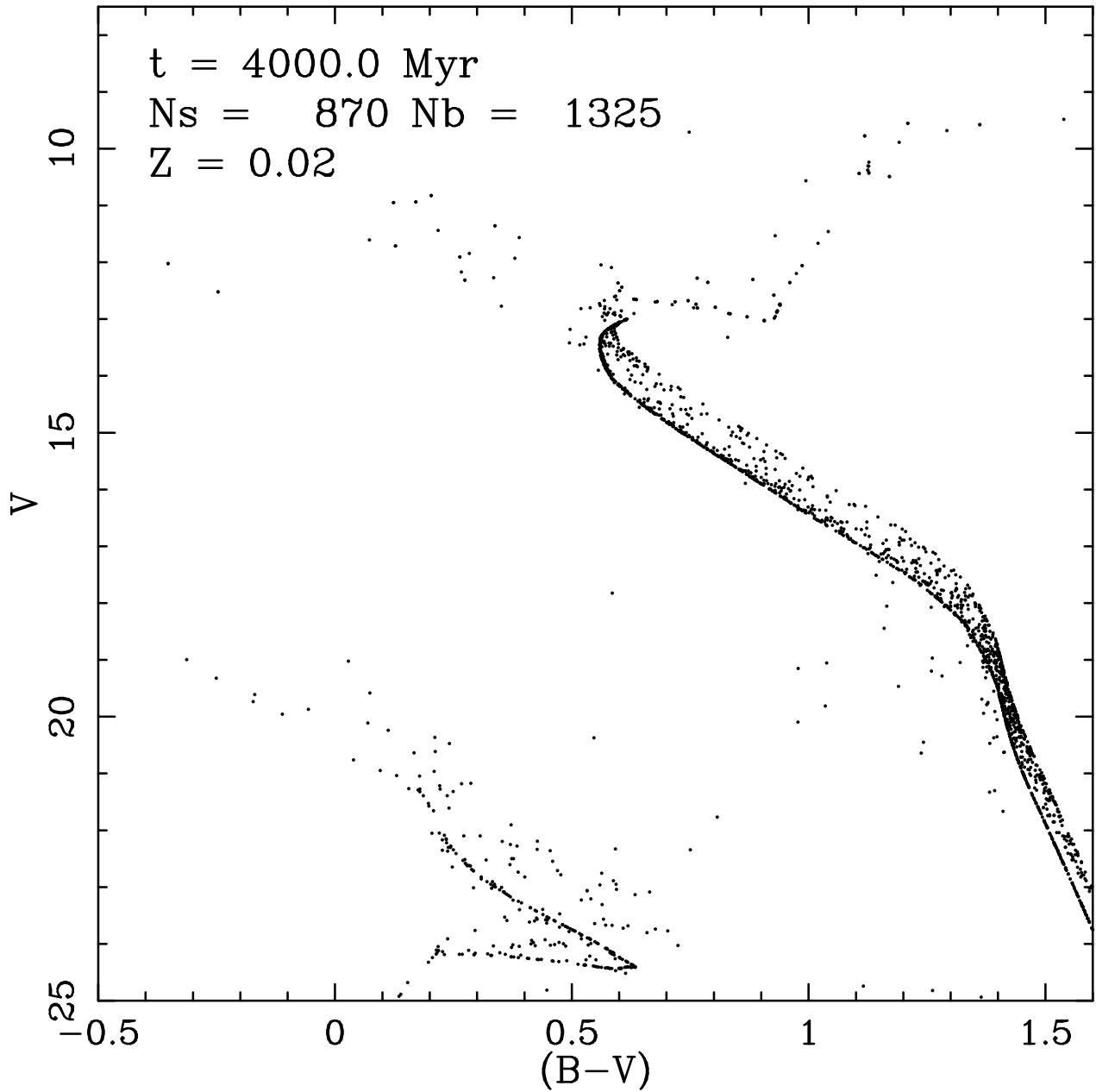}
\caption{
Cluster colour-magnitude diagram at $4.0\,$Gyr. 
Note that all binaries are assumed to be unresolved. 
To convert the luminosities and effective temperatures 
to magnitude and colour we have used the 
bolometric corrections given by Kurucz (1992) and, in the case of WDs, 
Bergeron, Wesemael \& Beauchamp (1995). 
A distance modulus of $9.7$ (Hurley et al. 2001) has been assumed to place the 
simulated cluster at the distance of M67. 
\label{f:fig10}}
\end{figure*}

\begin{figure*}
\includegraphics[width=168mm]{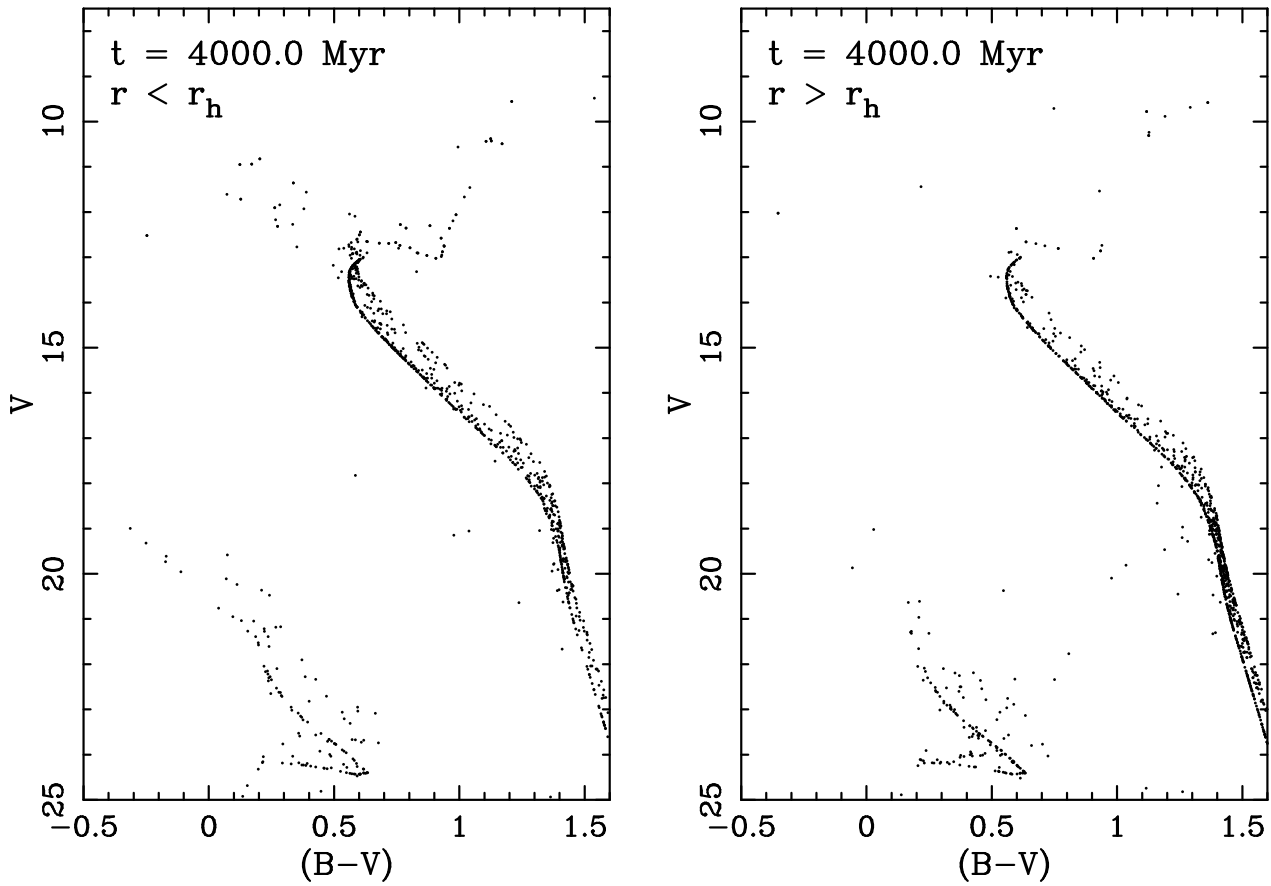}
\caption{
As for Figure~\ref{f:fig10} but for stars within the 
half-mass radius ($3.8\,$pc) of the cluster (left panel) and exterior to 
the half-mass radius (right panel). 
\label{f:fig11}}
\end{figure*}

\begin{figure}
\includegraphics[width=84mm]{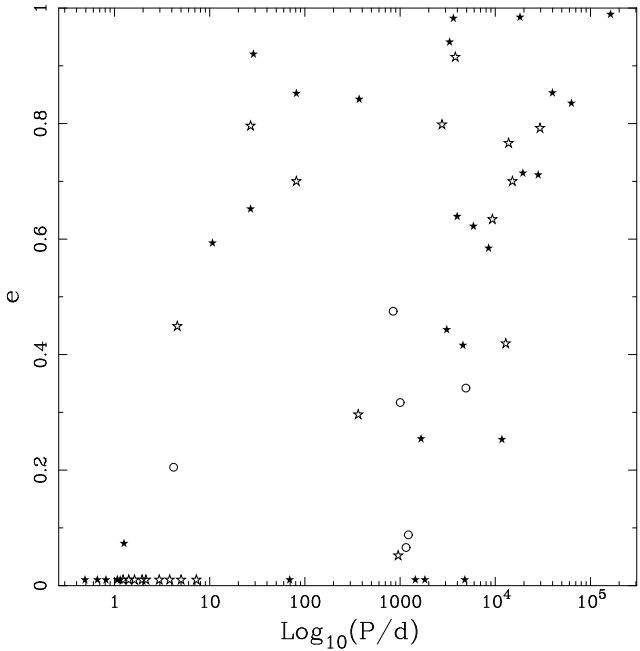}
\caption{
Distribution of periods and eccentricities for binaries found to contain  
a BS that were present in the M67 simulation (Model~2) at an age of 
$2\,$Gyr or later. 
The solid stars represent the orbital periods at the time of formation, 
when one of the stars became a BS or when a BS was exchanged 
in to a new binary (31 points). 
If the orbital parameters of any of these binaries subsequently experienced a 
significant change ($\Delta e > 0.05$ and/or $\Delta \log P / {\rm d} > 0.1$), 
owing to binary evolution or a perturbation, this is represented by an open 
star (21 points). 
Orbital parameters for the six BS binaries known to reside in M67 are denoted 
by open circles. 
\label{f:fig12}}
\end{figure}

\begin{figure}
\includegraphics[width=84mm]{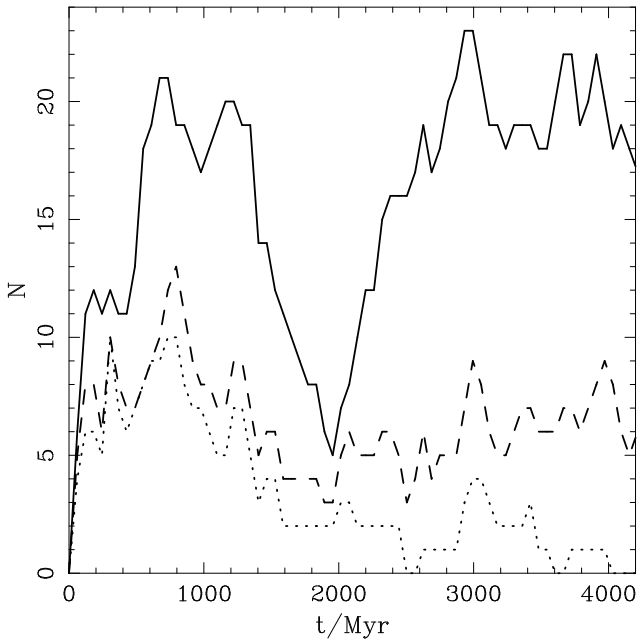}
\caption{
Number of BSs (solid line), BS-binaries (dashed line) and circular BS-binaries 
with $P < 100\,$d (dotted line) during the M67 simulation (Model~2). 
\label{f:fig13}}
\end{figure}

We have seen that formation scenarios for all of the various BSs 
observed in M67 exist in Model~2. 
The model at $4\,$Gyr has a binary fraction of about 0.5 for the BSs 
which is close to the fraction of 0.6 observed. 
Also, the ratio of BSs to the number of MS stars within two magnitudes 
of the MS turn-off ($13 < V < 15$) is 0.18 which matches well with the 
observed value of 0.14 (Ahumada \& Lapasset 1995). 
In fact this indicates that the model is over-producing BSs although if we 
look at raw numbers the model has 20 BSs at $4\,$Gyr compared to 
28 for M67 (Hurley et al. 2001). 
However, if we look at the CMD of M67 (e.g. Fig.~2 of Hurley et al. 2001; 
Montgomery, Marschall \& Janes 1993) the BSs appear to form two 
distinct groups, an obvious group of 10 BSs with $V < 12$ and the remainder 
much closer to the MS turn-off position. 
Some of these less obvious BSs actually lie below the MS turn-off by almost a 
magnitude. 
The observed BSs are identified by their position in the CMD whereas in the 
$N$-body model we have much more information about the stars and 
have used mass as the determining factor. 
Inspection of Figure~\ref{f:fig10} and comparison with the positions of the 
observed M67 BSs reveals that there are at least five stars near the MS 
turn-off of the model that we may have classified as BSs if using CMD 
position as the determining factor. 
Counting these stars as BSs would give good agreement of raw BS numbers 
for the model and M67. 
Also, two of the M67 BSs have proper motion membership probabilities 
of less than 80 per cent (Girard et al. 1989). 

We now focus on a comparison with the group of ten M67 BSs 
with $V < 12$. 
This sample has been well studied by Milone \& Latham (1992) revealing 
that six are single and four are in binaries. 
The orbital parameters of the BS-binaries are $P = 4.2\,$d and $e = 0.2$, 
$P = 846\,$d and $e = 0.5$, $P = 1\,003\,$d and $e = 0.3$, and 
$P = 1\,221\,$d and $e = 0.1$. 
The simulated cluster at $4\,$Gyr also has a pronounced group of 15 BSs 
(those with $|B-V| < 0.45$ in Figure~\ref{f:fig10}). 
Eight of these are single and seven are in binaries. 
If we further restrict this sample to $V < 12$ then we have eleven BSs 
with six in binaries. 
The widest of these six (\#1613 and \#3835) may not be detected as having 
a companion if observed so the mix of BSs in this sample could easily 
switch to 7 single BSs and 4 binary BSs. 
Either way, considering the stochastic nature of BS formation, the number 
of BSs in the restricted sample is a good match to that of the observed 
group of BSs as is the ratio of single BSs to binary BSs and the 
orbital combinations of the binary BSs.

\subsubsection{X-ray sources}

X-ray observations of old open clusters (van den Berg et al. 2004) and 
globular clusters (Pooley et al. 2003) have proved to be very efficient 
at detecting short-period interacting binaries. 
For open clusters there is a well-defined connection between age and 
X-ray activity (Randich 1997) where the latter declines with age owing 
to the spin-down of late-type stars which are initially rapid rotators. 
Therefore, X-ray detections in a cluster such as M67 are expected 
to be associated with stars that have been spun up in some way, 
typically by residing in a short-period binary where the spin period of 
the star is kept synchronized with the orbital period by tidal forces. 
An example of such a binary is an RS CVn system which contains a subgiant 
and a MS star companion in a short-period orbit. 
BY Draconis binaries are chromospherically active systems containing 
a late-type (spectral type F or later) MS star primary with a MS star 
companion and may also explain X-ray activity in old populations.  

A Chandra observation of M67 was recently reported by 
van den Berg et al. (2004). 
They detected 158 X-ray sources 
(with a limiting flux corresponding to an X-ray luminosity of 
$L_X \approx 10^{28} \, {\rm erg} \, {\rm s}^{-1}$). 
Optical counterparts that are proper-motion members of M67 were 
found for 25 sources and a further 12 sources had optical counterparts 
that are believed to be cluster members based on their position near 
the lower end of the M67 main sequence. 
Ten of the proper-motion members are binaries with periods 
less than $12\,$d and contain subgiant or MS stars.  
Approximately six of these appear to be classical RS CVn binaries 
of which three were detected earlier by Belloni , Verbunt \& Mathieu (1998).  
The twelve sources near the lower end of the MS are possible 
BY Draconis binaries.  

Table~\ref{t:table6} lists the parameters of the six RS CVn binaries 
observed in our M67 model at $4\,$Gyr. 
Five of these are the result of standard primordial binary evolution 
while one (\#1568) was formed in an exchange interaction. 
The number is a good match to the M67 observations although 
possibly we should not count the system in which the primary 
is filling its Roche lobe (\#2633) or the system with a period of 
$20\,$d (\#2383) when making the comparison as these fall outside 
of the classical RS CVn period range as defined by Hall (1976). 

\begin{figure}
\includegraphics[width=84mm]{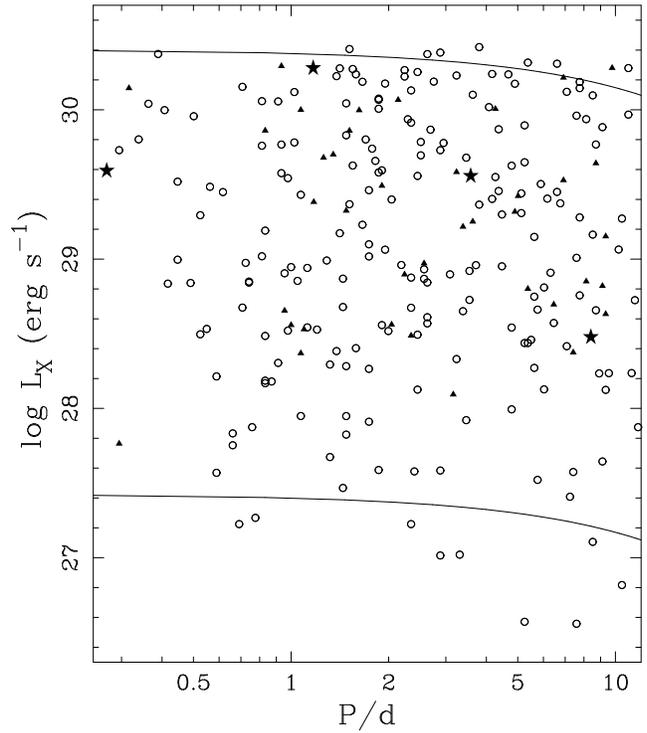}
\caption{
X-ray luminosity as a function of orbital period for MS-MS binaries 
with a primary mass of $1.0 M_\odot$ or less (open circles) and 
for MS-WD binaries (solid triangles). 
The upper solid line shows the X-ray luminosity for a $1.0 M_\odot$ MS 
star at an age of $4\,$Gyr calculated from Equation~2 as a function 
of orbital period, assuming that the star is in a binary of that period and 
experiencing synchronous rotation. 
The lower solid line is for a $0.1 M_\odot$ MS star. 
We also include data points (solid stars) for the possible BY Draconis systems 
identified by van den Berg et al. (2004) for which an orbital period is known 
($L_X$ values for these systems supplied by M. van den Berg, private 
communication). 
\label{f:fig14}}
\end{figure}

To investigate the incidence of BY Draconis binaries we look at 
all MS-MS binaries where the primary is F-type or later 
($M_1 \leq 1.0 M_\odot$) and the orbital period is $12\,$d or less. 
This cutoff in period is used in observational work 
(e.g. van den Berg et al. 2004) and is thought to ensure that 
all binaries in the sample are synchronously 
rotating at the age of M67. 
Using the relation between X-ray luminosity and rotation rate, $\Omega$, 
proposed by Walter (1982), 
\begin{equation}
\log L_X/L_{\rm bol} = -3.14 - 0.16 / \Omega \, , 
\end{equation} 
where $L_{\rm bol}$ is the bolometric luminosity of a star and $\Omega$ 
is in units of rotations/day, we can estimate an X-ray luminosity for each of the 
BY Draconis binaries. 
These are plotted in Figure~\ref{f:fig14} as a function of orbital period. 
The upper limit to the X-ray luminosity at any period is given by a $1.0 M_\odot$ star 
in synchronous rotation with its orbit -- to emphasize this we have plotted the 
behaviour of Equation~2 for such a situation (upper solid line in Figure~\ref{f:fig14}). 
Correspondingly the lower limit occurs for a $0.1 M_\odot$ star -- the lowest 
mass of star that we considered (see the lower solid line in Figure~\ref{f:fig14}). 
A small number of systems in our M67 model have $L_X$ less than this lower limit 
and these are binaries where both stars are less massive than $0.2 M_\odot$ and 
synchronous rotation is yet to be achieved. 
We have also included data points for the BY Draconis systems identified by 
van den Berg et al. (2004) for which orbital periods are known 
(four of twelve). 
We find 172 MS-MS systems with $L_X > 10^{28} \, {\rm erg} \, {\rm s}^{-1}$. 
Also shown in Figure~\ref{f:fig14} are 19 short-period MS-WD binaries 
which have X-ray luminosities in this range and for which the WD component 
has cooled sufficiently that the binary appears near the MS in the CMD. 

We have an over-abundance of potential MS X-ray sources compared to 
the Chandra detections for M67. 
However, the Chandra pointing involved six CCD detectors with each having 
an area of $8.4' \times 8.4'$ so that the coverage was only about 1/30th of the 
full area of M67 (for a tidal radius of $60'$). 
Also, the sensitivity of the Chandra observations decreases towards the edge 
of the detector and the limiting luminosity will be greater than 
$10^{28} \, {\rm erg} \, {\rm s}^{-1}$ for the outer detectors 
(M. van den Berg, private communication). 
The centre of the cluster was covered by Chandra so we do not expect the 
RS CVn sample to suffer greatly from incompleteness as these are 
relatively massive binaries but 
BY Draconis binaries will not necessarily reside in the 
central regions of the cluster, especially those with cool primaries. 
Therefore there may be a substantial population of these systems yet to 
be observed.

\subsubsection{White dwarfs}

Hurley \& Shara (2003) used $N$-body models to conduct a detailed 
investigation of the behaviour and appearance of the white dwarf 
population in dense star clusters. 
They found that the mass fraction of WDs can be significantly enhanced 
by the dynamical evolution of the cluster -- by as much as a factor of 2  
in old open clusters. 
This enhancement is relative to the mass fraction expected if 
the same populations were evolved with 
basic population synthesis and with the same stellar and binary 
evolution algorithms used in the $N$-body code. 
A combination of mass segregation and the presence of a tidal field 
mean that, as a cluster evolves, low-mass MS stars are preferentially 
stripped from the cluster and the WD mass fraction diverges from that 
of the field population. 
We find the same behaviour in our M67 model. 
The expected WD mass fraction at $4\,$Gyr is $f_{\rm WD} = 0.1$ when 
we perform population synthesis on the initial population. 
However, the M67 $N$-body model has $f_{\rm WD} = 0.15$ at $4\,$Gyr. 
The average mass of the cluster WDs starts high and then decreases with 
time as progressively less-massive WDs are born. 
After $1\,$Gyr the WD average mass is $0.82 M_\odot$ and after $4\,$Gyr 
it is $0.64 M_\odot$. 
On the other hand the average mass of MS stars is $0.52 M_\odot$ for 
the starting model and initially decreases owing to evolution of the 
massive MS stars followed by an increase to $0.56 M_\odot$ at $4\,$Gyr 
because of tidal stripping. 
In other words, the average mass of MS stars is nearly constant. 
Based on these values we would expect the WDs to be preferentially found 
near the centre of the cluster as a result of mass segregation. 
This expectation is strengthened when we consider that WDs are born 
from stars that were previously the most massive MS stars in the cluster. 
What we observe in our M67 model is that the half-mass radius of the 
WDs is $0.64\,$pc which is indeed centrally concentrated when compared 
to a half-mass radius of $3.8\,$pc for all stars. 
If we look at only single WDs the half-mass radius increases to $1.28\,$pc 
(for comparison the half-mass radius of the cluster binaries is $0.37\,$pc). 
The upshot of all this is that low-mass MS stars are more likely to reside 
in the outer regions of the cluster and are thus more vulnerable to 
escape from the cluster than WDs. 

\begin{figure}
\includegraphics[width=84mm]{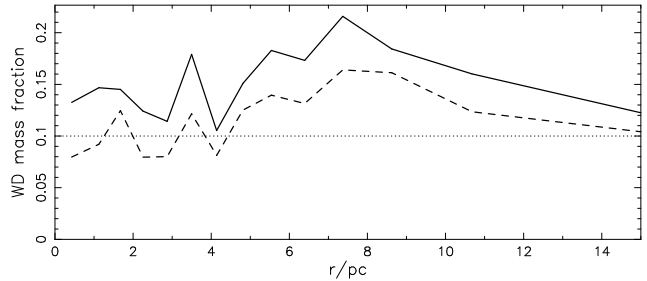}
\caption{
Mass fraction of white dwarfs in Model~2 at $4\,$Gyr as a function of 
radius. 
The radial bins are chosen so that 150 stars are sampled in each bin. 
Results for all WDs (solid line) and for only single WDs and double WDs 
(dashed line) are shown. 
The dotted line at $f_{\rm WD} = 0.1$ is the value expected from isolated 
population synthesis of the initial stars. 
\label{f:fig15}}
\end{figure}

Figure~\ref{f:fig15} shows the variation of WD mass fraction as a function 
of radial position in the M67 model. 
We see that the value of $f_{\rm WD}$ depends strongly on which portion 
of the cluster is observed. 
Somewhat surprisingly the greatest enhancement compared to the 
population synthesis value is found in the 6 to $10\,$pc region, exterior 
to the cluster half-mass radius. 
There is less but noticeable enhancement in the core which should contain 
a significant population of massive old WDs 
as well as double WDs and less-massive young WDs. 
However, the central regions are also dominated by massive stars and binaries of 
all types which reduces $f_{\rm WD}$. 
This also means that young WDs move outwards from the centre  
as they cool. 
Outside the half-mass radius there is a lack of massive stars and binaries, 
as well as young WDs (see also Figure~\ref{f:fig11}). 
This increases the relative mass of the WD population. 
There is one radial bin near the half-mass radius where $f_{\rm WD}$ is similar 
to the population synthesis value. 
Unfortunately the behaviour of $f_{\rm WD}$ in the vicinity of this region is 
rather erratic and this makes it difficult to suggest that observations to 
determine $f_{\rm WD}$ should be conducted near the half-mass radius. 
In fact our results indicate that M67 is dynamically old enough that a 
measurement of $f_{\rm WD}$ cannot be used to yield information 
about the IMF. 
Also shown in Figure~\ref{f:fig15} is the value of $f_{\rm WD}$ calculated 
if we ignore WDs in binaries with non-WD companions. 
So this value corresponds to single WDs and double WDs found on or 
near the WD sequence in the cluster CMD which are the WDs most likely 
to be observed in a real cluster. 

Richer et al. (1998) derive a WD mass fraction of 0.09 for M67. 
This is based on finding 85 WDs down to the termination point of the 
WD cooling sequence (their deep survey reached $V = 25$). 
Correcting this number for WDs hidden in binaries and taking a  
50 per cent binary fraction they estimate that there could be as many as 
150 WDs in M67. 
Assuming an average mass of $0.7 M_\odot$ for the WDs and 
using a calculated total mass of $1\,080 M_\odot$ for M67 they 
arrive at the quoted value of $f_{\rm WD}$. 
Richer et al. (1998) note that this number appears to be low when 
compared to the number of giants they observed.  
Based on a population of 87 giants they expected about $60$ WDs 
with a cooling age less than $1\,$Gyr but found only 24. 
Our M67 model contains 226 single WDs, 60 double-WD binaries 
and 145 WDs in binaries with a non-WD companion. 
Of the single WDs 54 have a cooling age less than $1\,$Gyr. 
There are 22 double WDs with at least one bright component and 
33 bright WDs contained in other binaries (some of these 
with low-mass MS companions may appear near the WD sequence). 
The model also contains 59 giants which gives an approximate 
1:1 ratio of bright WDs to giants if we consider only the single WDs. 
If our model is to be believed this suggests that the observations 
of WDs in M67 are incomplete. 

Nine double-WD binaries with a total mass in excess of the Chandrasekhar 
mass ($M_{\rm Ch} = 1.44 M_\odot$) and merger timescales owing to 
gravitational radiation of less than the age of the Galaxy (about $13\,$Gyr) 
are produced in Model~2. 
One of these contains two oxygen-neon (ONe) WDs and three have an 
ONe and a carbon-oxygen (CO) WD component. 
The outcome of WD-WD mergers involving an ONe WD is believed to 
be an accretion-induced collapse (AIC: Nomoto \& Kondo 1991) to form 
a neutron star remnant. 
This is the assumed outcome when such an event occurs in an 
{\tt NBODY4} simulation. 
Mergers of CO-CO WD binaries with $M > M_{\rm Ch}$ are possible 
causes of Type Ia supernovae (Yungelson \& Livio 2000) but may 
instead lead to AIC and the formation of a neutron star (Saio \& Nomoto 1998). 
Shara \& Hurley (2002) found that the incidence of these possible Type Ia events 
increased by as much as a factor of 10 in the environment of an open cluster. 
This result was based on simulations of $20\,000$ stars with a 10 per cent primordial 
binary population and the Eggleton, Fitchett \& Tout (1989) orbital separation 
distribution used to set up the binaries. 
From the $12\,000$ primordial binaries in our simulation, population synthesis 
with the BSE code predicts that we should get three super-Chandrasekhar 
CO-CO WD mergers. 
So the five that we did find represents an enhancement but not of the order 
found by Shara \& Hurley (2002). 
However, it is too early to tell if the degree of enhancement depends on 
primordial binary fraction.  
The two additional CO-CO mergers in Model~2 are the result of perturbations 
to the orbits of primordial binaries that brought the component stars closer together. 
In each case a CO-CO WD binary would have formed regardless of the 
perturbation but the orbital period would have been greater and merging 
would not have occurred. 

At $4\,$Gyr one of the ONe--CO WD merger candidates remains in the cluster. 
It is not expected to merge for another $8\,$Gyr which is after the 
cluster will have completely dissolved. 
The other two ONe--CO WD binaries merged and formed neutron stars 
at ages of $63$ and $75\,$Myr. 
The ONe--ONe WD binary formed after $61\,$Myr and the components 
were expected to merge shortly afterwards but in the interim the binary 
was ejected from the cluster. 
The five CO--CO WD binaries merged prior 
to $4\,$Gyr, the first after $110\,$Myr of evolution and the last at 
$2\,666\,$Myr. 
Also formed during the simulation were ten double helium WD binaries 
with merging timescales less than $13\,$Gyr. 
Four of these escaped from the cluster prior to $4\,$Gyr. 
The other six remain in the cluster with periods of $0.04$ to $0.08\,$d 
and three are experiencing steady mass transfer. 
In the BSE and {\tt NBODY4} codes mass transfer from one WD to another 
occurs on a dynamical timescale if the mass ratio (donor/accretor) 
exceeds 0.628 and the outcome is coalescence of the WDs. 
If two He WDs merge it is assumed that the temperature produced 
is enough to ignite the triple-$\alpha$ reaction and that the nuclear energy 
released destroys the star. 
Four of the ten He--He WD binaries have mass ratios that satisfy this condition 
and two of them remain in the cluster at $4\,$Gyr. 
Iben (1990) discusses the possibility of subdwarf-O or B stars forming 
from merged He WDs which ignited helium at 
the base of the accretion layer. 
This alternative scenario of helium star formation had previously been 
mentioned by Webbink (1984).

\subsection{Luminosity Functions}
\label{s:res_sub3}

\begin{figure}
\includegraphics[width=84mm]{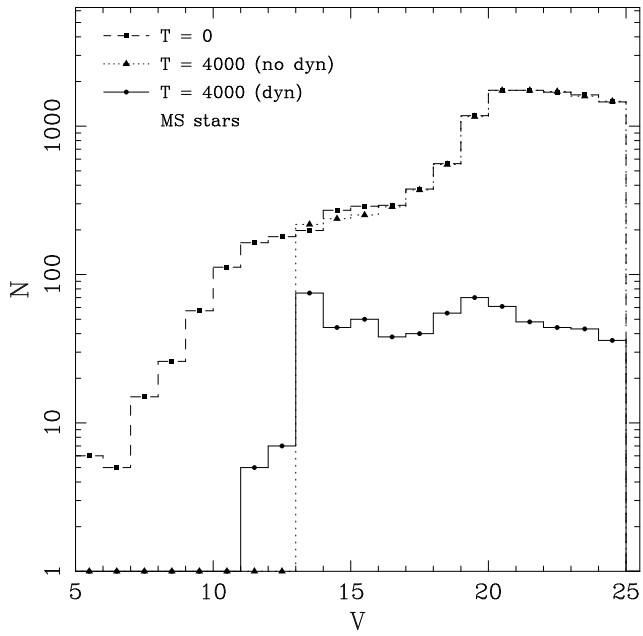}
\caption{
Luminosity function of single main-sequence stars in Model~2 at time zero 
(dashed line, $12\,000$ stars) and at $4\,$Gyr (solid line, $616$ stars). 
Also shown is the luminosity function for the initial stars evolved to 
$4\,$Gyr with the population synthesis code (dotted line, $11\,341$ stars). 
The histograms are not normalized. 
\label{f:fig16}}
\end{figure}

The luminosity function (LF) for the 616 single MS stars in Model~2 at $4\,$Gyr 
is shown in Figure~\ref{f:fig16}. 
This is compared to the LF of the initial model ($12\,000$ single MS stars) 
and also the LF at $4\,$Gyr for the model evolved with the population 
synthesis code. 
The latter distribution we call the non-dynamical LF and it contains 
$11\,341$ stars up to the MS turn-off at $V = 13$. 
So 659 stars have evolved off the MS by $4\,$Gyr. 
The primordial and non-dynamical MS LFs are identical for $V > 16$ 
and very similar up to $V = 13$ apart from some fluctuation 
owing to the evolution of near-turn-off stars. 
So the non-dynamical MS LF at $4\,$Gyr can be used to infer the IMF 
of the population. 
If the slope of the cluster MS LF matches that of the non-dynamical LF 
it would also retain an imprint of the IMF. 
However, we can see from Figure~\ref{f:fig16} that the two are far from 
being a good match. 
The cluster LF has been significantly flattened by the preferential 
evaporation of low-mass stars from the cluster and by merging 
within binaries creating new MS stars near the turn-off. 

\begin{figure*}
\includegraphics[width=168mm]{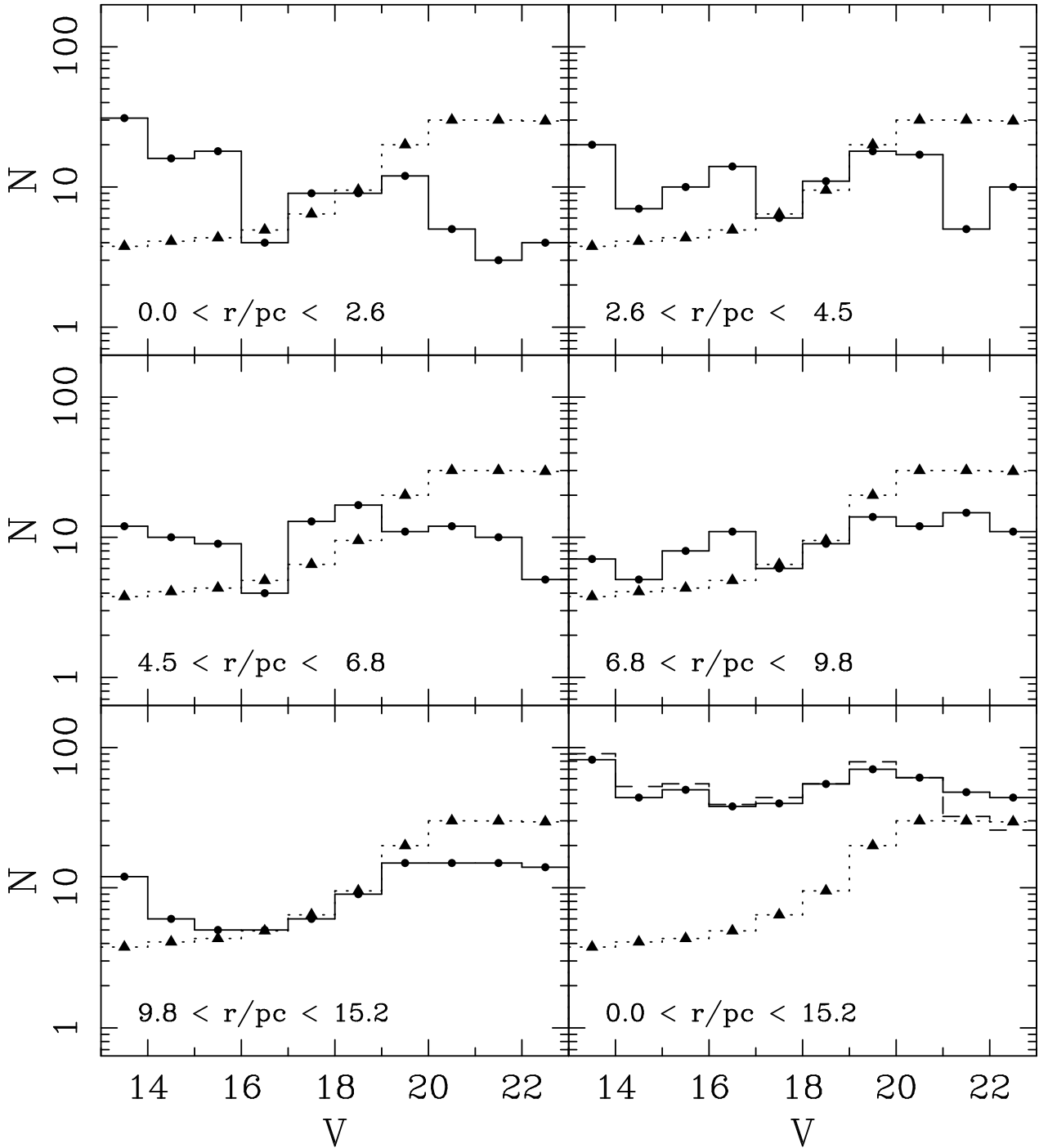}
\caption{
Luminosity function of single main-sequence stars for $13 < V < 23$ 
in Model~2 at $4\,$Gyr split into five radial regions. 
The regions are chosen so that there is a similar number of stars 
in each (about 106). 
The sixth panel (lower-right) shows the combined distribution. 
We compare this with the luminosity function for single MS  
stars and MS-MS binaries with $q < 0.5$ (dashed line). 
Also shown (dotted line with solid triangles) in each panel is the 
population synthesis MS luminosity function at $4\,$Gyr 
but with the number of stars in each bin reduced by a factor 
of 60 in order to aid comparison of the distribution slopes. 
\label{f:fig17}}
\end{figure*}

In Figure~\ref{f:fig17} we split the cluster into five radial regions to 
examine the radial dependence of the MS LF slope and 
investigate whether there is any region which retains sufficient 
information about the IMF. 
There is a clear radial dependence with the central region containing 
many more massive stars than low-mass stars and the slope 
becomes flatter as we move outwards through the cluster. 
The normalized non-dynamical MS LF is compared to the cluster LF 
in each region and we can see that it is only in the outer regions 
that the slopes converge but, even then, the 
cluster LF is deficient in very low-mass stars and has an over-abundance 
of stars near the turn-off. 
Terlevich (1987) pointed out that in non-isolated clusters, heating by 
the galactic tide has the effect 
of making the velocity distribution isotropic and as such orbits ejected 
from the centre will tend to avoid returning there. 
This effect is in addition to mass segregation and helps in understanding 
the presence of massive MS stars in the outer regions. 
The LF for the five regions combined is also shown in Figure~\ref{f:fig17} 
and is essentially flat with a clear preference for massive MS stars 
compared to the non-dynamical (or field) distribution. 
Fan et al. (1996) show that the overall MS LF for M67 is quite flat and that 
the centre of the cluster is deficient in low-mass stars. 
They also found a radial dependence for the mass function slope. 
Bonatto \& Bica (2003) confirm the tendency for massive stars in M67 
to be more centrally concentrated  and that the central regions of the 
cluster show a turnover in the cluster luminosity function at low masses. 
They find that the halo of the cluster is enriched in low-mass stars and 
their MS LF for the outer regions of M67 gives the best match to the 
expected LF slope for field stars (Kroupa, Tout \& Gilmore 1993). 
 
We have presented LFs of single MS stars as this is the ideal 
situation to deal with if one wants to make inferences about the IMF 
of the stellar population. 
Evolved stars such as giants complicate matters because they suffer 
mass loss and their luminosity evolution is not as well understood, and 
binary star evolution is obviously more uncertain than that of single stars. 
When dealing with observed data the goal is the same but the process 
is not as straightforward. 
Removing evolved stars from the LF by inspection of the CMD is not too 
difficult. 
Binaries with high mass-ratios can also be removed using the method of 
CMD inspection. 
However, low mass ratio binaries are problematic and will lead to contamination 
of the LF. 
In Figure~\ref{f:fig17} (sixth panel) we compare the slope of the LF for single MS 
stars to that of the same LF but with binaries with mass ratios less than 0.5 
included as well. 
There are 532 single MS stars and 380 low--q binaries and the distributions 
have been normalized so that the slopes may be compared. 
We see that the inclusion of the low--q binaries does not affect the LF slope 
except at the low-mass end where binaries are biased towards high--q. 
Therefore, our model indicates that contamination of the MS LF by binaries 
is not necessarily a problem when using the shape of the distribution to 
infer the slope of the IMF.

\section{Discussion and Summary}
\label{s:discus}

A conclusion to draw from inspection of our two binary rich open 
cluster models is that in an open cluster, at least, one requires a substantial 
population of seed binaries capable of making BSs in order to explain the observed 
numbers of BSs. 
For Model~1 we used the birth period distribution suggested by Kroupa (1995b) 
to generate the orbital parameters of the $9\,000$ primordial binaries. 
Binary population synthesis told us to expect these binaries to produce just one BS 
after $4\,$Gyr of evolution and this would be via Case~C mass transfer in a wide binary. 
After performing the $N$-body simulation we indeed found one BS in the model at $4\,$Gyr. 
For Model~2 we included $12\,000$ primordial binaries and chose the initial 
separation of each binary from a flat logarithmic distribution with an upper limit of $50\,$au. 
In this case we expected 25 BSs at $4\,$Gyr with 75 per cent being produced 
from Case~A mass transfer in short-period systems. 
What we found after $4\,$Gyr of cluster evolution was 20 BSs with seven of 
these produced from non-perturbed primordial binaries. 
So the expected BSs were depleted while others were created. 
A major difference between the primordial binary setups of Models~1 and 2 is 
that the frequency of short-period binaries is much less in Model~1 and no 
BSs are expected to be produced from Case~A mass-transfer in these binaries. 
The formation rate of non-primordial short-period binaries is very small in 
open cluster simulations and when it does occur it is generally as a result of 
a binary-binary encounter involving at least one existing short-period binary. 
Also, in the moderate density conditions of an open cluster we do not see BS 
formation from direct collisions of single stars while relatively wide primordial 
binaries may be destroyed in encounters with other stars before reaching a 
Case~B or C mass-transfer stage. 
These factors explain the absence of BSs in Model~1 and tell us that the 
cluster environment on its own is not efficient at producing BSs. 
It is important to have a population of short-period binaries that either 
form BSs directly or become involved in dynamical encounters with other 
cluster stars and binaries and subsequently produce BSs. 
Our results indicate that period distributions such as Kroupa (1995b) and 
Duquennoy \& Mayor (1991) that predict minimal BS formation from the 
standard Case~A mass-transfer scenario are not suitable as initial conditions for 
open cluster binaries. 

In Model~2 we observed formation pathways to explain the full variety of BSs 
found in M67, single BSs, super-BSs and binary BSs with a range of 
period--eccentricity combinations. 
At $4\,$Gyr we found that 50 per cent of the BSs are single and, of those in 
binaries, all but one has an eccentric orbit. 
This is in contrast to the binary population synthesis expectation for the primordial 
binaries of the model which gives a mix of 75 per cent single and only circular binaries. 
This tells us that dynamical encounters within the cluster environment play 
an important role in defining the nature of the BS population but perhaps not in boosting 
actual BS numbers. 
Less than half of the BSs were formed from primordial binaries that did not have 
their evolution altered in any way by dynamical encounters and the majority of 
these involved Case~A mass transfer followed by coalescence to form a single BS. 
Other cases involved perturbations to the orbits of primordial binaries that induced 
mass transfer or, in some cases, a delay of mass-transfer after a primordial binary 
became involved in a short-lived exchange encounter and emerged intact but 
with its orbital parameters altered. 
Mass transfer also led to BS formation in binaries created from exchange interactions. 
The alternative to BS formation via mass transfer is the collision of two MS stars 
at pericentre in a highly eccentric binary. 
This was observed to occur in primordial binaries with mildly eccentric orbits that had 
their eccentricity increased by perturbations received from nearby stars or binaries, 
often after becoming involved in a three- or four-body hierarchy. 
Exchange interaction also created highly eccentric binaries in which the 
component stars eventually collided to form a BS. 
Mapelli et al. (2004) showed that the radial distribution of BSs observed in the 
globular cluster 47 Tucanae is best explained by formation from primordial 
binaries (mass-transfer scenario) in the outer regions of the cluster and a mixture 
of exchange-induced collisions and primordial binary evolution in the core. 
So even though the stellar environment is very different in the core of 47 Tucanae 
than in M67 it seems that a mixture of BS formation scenarios is favourable in both. 
Davies, Piotto \& De Angeli (2004) demonstrated that exchange interactions can be 
detrimental to BS formation via primordial binaries in massive globular clusters. 
They consider that exchanges produce binaries with increasingly more-massive 
components and this could lead to binaries that would have created a BS to be 
observed in a globular cluster now being replaced by binaries that form BSs much 
earlier. 
Their experiments showed that an average number of encounters per binary of 
about 5 to 10  was beneficial for BS production and that anything in excess of 
this would seriously hamper the BS formation rate. 
This result is tuned to the age and turn-off mass of globular clusters 
and is also dependent on the distribution of binary separations but we note 
that in our open cluster simulations only a few per cent of the binaries were 
involved in multiple exchange encounters. 

Because a significant fraction of the BSs in our model are influenced 
in some way by the formation of triple and quadruple subsystems it would 
seem prudent to consider including a primordial population of these. 
Especially as triple and higher-order systems are observed in young 
open clusters (Schertl et al. 2003) and the field (Tokovinin 1997). 
Sandquist (2004) mentions nine possible triple systems in M67. 
Also, van den Berg et al. (2001) identified the BS S1082 in M67 
(classified as single by Milone \& Latham 1992) with a spectroscopic 
close binary and based on X-ray emission it is now thought to be 
a triple system containing an RS~CVn star with the BS as the outer 
component (van den Berg et al. 2004). 
Our M67 model at $4\,$Gyr contains 20 hierarchical triples and 
there are 20 to 30 stable triples in the Model~2 simulation at any time. 
None of the BSs in the model at $4\,$Gyr are in triple systems although 
one BS-binary is loosely bound to another binary with an orbital 
separation of $5 \times 10^4\,$au. 
There are instances of triples containing a BS at other times, primarily as 
a result of a BS forming in an eccentric binary collision within a four-body 
or higher order system and remaining bound to other members of the 
system. 
These are not generally long-lived. 
Short-lived triple systems are also found to be responsible for inducing an eccentricity 
in the orbits of previously circular BS binaries. 
This is the Kozai effect (Kozai 1962) produced by a cyclic relation between 
the inner eccentricity and the orbital inclination and which is modelled 
in {\tt NBODY4} (Aarseth 2003). 
One of the RS CVn binaries in Model~2 at $4\,$Gyr (\#1568) is the inner binary 
of a triple system with a $1.3 M_\odot$ MS star and an outer period of about $1\,000\,$d. 
The capability to include primordial triple and quadruple systems has recently 
been added to {\tt NBODY4} and we plan to utilize this in future simulations. 

Mathieu et al. (2003) report the finding of two spectroscopic binaries in M67 
with a position in the CMD about $1\,$mag below the subgiant branch. 
They are high-probability 
proper-motion members and have also been confirmed as X-ray sources 
(van den Berg, Verbunt \& Mathieu 1999). 
Based on their CMD position the primaries of these binaries are termed 
sub-subgiants and it has been postulated that they are the products of stellar 
encounters on non-standard evolutionary tracks (Mathieu et al. 2003). 
One binary is circular with a period of $2.82\,$d and the other has 
an eccentricity of 0.21 and a period of $18.4\,$d. 
The eccentric binary is in the core of M67 and shows high reddening which implies 
a subgiant with extinction (Mathieu et al. 2003). 
The star lying below the subgiant branch in our M67 model was formed 
in a CE event that saw $0.29 M_\odot$ of material 
ejected from the star when the cores merged. 
Extinction from this material was not considered when calculating the CMD 
position of the star but this possibility and its position below the subgiant 
branch make it a candidate sub-subgiant explanation. 
However, to explain the binary nature the star would need to have been a member 
of a triple system in which the merged star remained bound to the third component. 
This is not an unrealistic scenario and the inclusion of primordial triples would 
make it more likely. 

The M67 model we have presented here is a marked improvement on the 
model previously reported in Hurley et al. (2001). 
A major reason for this is in the construction of the models. 
The current model is the 
result of an $N$-body simulation that was evolved from zero-age whereas the 
Hurley et al. (2001) $N$-body model started at $2.5\,$Gyr. 
Also Hurley et al. (2001) subjected their model to an unusually strong tidal field 
in an attempt to produce a cluster that contained the observed mass of M67 
within a tidal radius of $10\,$pc. 
More recent observations suggest that the tidal radius of M67 is actually greater 
than this. 
To achieve our preferred M67 model (Model~2 at $4\,$Gyr) we did make an 
alteration to the tidal field. 
Model~2 placed the cluster on an orbit at $8\,$kpc 
from the Galactic centre as opposed to $8.5\,$kpc for the standard orbit used 
in Model~1. 
This is well within the bounds allowed by the actual orbit of M67 
and an orbital speed of $220\, {\rm km} \, {\rm s}^{-1}$ was used 
for both Models~1 and 2. 
Hurley et al. (2001) used $350\, {\rm km} \, {\rm s}^{-1}$. 
So the modification of the tidal field for Model~2 is nowhere near as 
extreme as for the Hurley et al. (2001) model. 
In fact, according to Baumgardt (1998) it is the tidal radius at perigalacticon that 
determines the cluster dissolution timescale and considering that M67 has a 
perigalacticon of $6.8\,$kpc we could have chosen an even stronger tidal field 
for Model~2, and thus started with a greater number of stars and binaries. 

Results are also better for our new M67 model because half of the BSs are 
found in binaries compared to only one out of 22 found by Hurley et al. (2001). 
The raw numbers of BSs are similar but the Hurley et al. (2001) model actually 
did better at creating BSs by dynamical means -- they expected only half as many 
BSs as their model produced whereas our model actually produces slightly less 
than expected. 
Comparison of the two models is not straightforward because different distributions 
were used for the primordial binary separations and other factors such as the 
tidal field used by Hurley et al. (2001) significantly alter the evolution of the 
cluster. 
Also, the semi-direct nature of the Hurley et al. (2001) model, which started 
with $5\,000$ single stars and $5\,000$ binaries at an age of $2.5\,$Gyr, 
complicates direct comparison. 
Our model with initially $12\,000$ single stars and $12\,000$ binaries 
actually contained $4\,322$ single stars and $4\,859$ binaries after $2.5\,$Gyr 
of evolution so the numbers are comparable at this stage. 
However, in an attempt to compensate for skipping $2.5\,$Gyr of $N$-body 
evolution Hurley et al. (2001) selected their primordial binaries using a 
mass function biased towards higher masses and then evolved these to 
an age of $2.5\,$Gyr with the BSE code. 
Comparison of this population with our $4\,859$ binaries at $2.5\,$Gyr shows 
that the Hurley et al. (2001) model has a higher proportion of binaries with a 
combined mass in excess of the cluster turn-off mass at $4\,$Gyr. 
The semi-direct model also evolved at higher central density 
($\rho_c > 1\,000\, {\rm stars} \, {\rm pc}^{-3}$ for the majority of the simulation) 
and with a smaller half-mass radius. 
This would explain the increased incidence of dynamical BS formation. 

It is reassuring that the structure of our M67 model provides a good match to 
the observed structure of M67. 
The half-mass radius, the tidal radius and the mass of the model and cluster are all 
in agreement. 
The surface density profile of the model is also able to reproduce the observed 
M67 profile, except in the core where the model has too many stars. 
We have constructed a surface brightness profile for the model using the same 
software as used for treating observed data and found that it is very well 
fitted by an Elson, Fall \& Freeman (1987) distribution. 
This distribution was determined from young and intermediate age clusters in 
the Large Magellanic Clouds which do not show the same degree of tidal 
truncation as the old globular clusters of our Galaxy to which the King (1962) 
models were fitted. 
Differences between the Elson, Fall \& Freeman (1987) model and the King (1962) 
model are only apparent near the tidal radius of a cluster. 
The luminosity functions for MS stars in various spatial regions of our M67 model 
also give good agreement with the observed behaviour. 
Our M67 model has done a very good job of simulating the 
blue straggler and RS CVn stars observed in M67. 
So the model places strong constraints on the parameters of the primordial 
binary population in this cluster. 
Also, by matching the properties of populations for which observations are likely 
to be complete (BSs and RS CVn stars), we can use the model to make 
predictions about the completeness of other populations (e.g. white dwarfs and 
BY Draconis stars). 

The process of tailoring a simulation to the parameters of a particular cluster and 
taking the time to compare the real and model data in a consistent manner has 
certainly proved fruitful. 
An effort along these lines was made previously by Portegies Zwart et al. (2001) 
who looked at young open clusters such as the Hyades using models of 
$N \simeq 3\,000$, whereas we have focussed on old clusters where the 
interaction between dynamical, stellar and binary evolution is more pronounced. 
We look forward to taking this approach with other rich open clusters of various ages,
such as NGC$\,2099$ (Kalirai et al. 2001) and 
NGC$\,188$ (Stetson, McClure \& VandenBerg 2004), 
in order to further understanding the initial conditions, dynamical history and stellar 
populations of these objects. 
It can then be taken into the globular cluster regime when the 
petaflops speed GRAPE-DR (Makino et al. 2003) eventually becomes available. 
We also urge those working on observations of star clusters to take full advantage 
of dynamical models when interpreting data.

\section*{Acknowledgments}

We thank Charles Bonatto for providing us with M67 
data and allowing this to be used in the paper. 
Thanks to Dougal Mackey for providing his surface brightness profile 
software as well as advice on how to use it. 
We are grateful to Maureen van den Berg for clarifying the M67 Chandra 
data and making useful comments on the manuscript. 
JRH thanks the Australian Research Council for a Fellowship 
and is grateful to the Institute of Astronomy, Cambridge for 
hosting a visit during this work. 
CAT thanks Churchill College for a Fellowship. 
We acknowledge the generous support of the Cordelia Corporation and
that of Edward Norton which has enabled AMNH to purchase 
GRAPE-6 boards and supporting hardware.

\clearpage

\clearpage

\clearpage

\clearpage

\clearpage 

\clearpage

\clearpage

\clearpage

\clearpage

\begin{table*}
\begin{minipage}{126mm}
\caption{
Parameters of the starting models at time zero for simulations performed 
in this work (see text for details). 
\label{t:table1}
}
\begin{tabular}{lrr}
\hline
 & Model~1 & Model~2 \\
\hline
    $N_{\rm s}$ & 9000 & 12000 \\
    $N_{\rm b}$ & 9000 & 12000 \\
    $R_{\rm G}$/kpc & 8.5 & 8.0 \\
    $v_{\rm G}$/km/s & 220 & 220 \\
    density profile & Plummer & Plummer \\
    binary periods & Kroupa & flatlog \\
    $M_0 / M_\odot$ &  14405 & 18687 \\ 
    $T_{{\rm rh},0}$/Myr & 300 & 290 \\      
    $R_{\rm t}$/pc & 34.4 & 31.8 \\
    $R_{\rm h}$/pc & 4.3 & 3.9 \\
\hline
\end{tabular}
\end{minipage}
\end{table*}

\begin{table*}
\begin{minipage}{126mm}
\caption{
General results at $4\,$Gyr for simulations performed in this work. 
\label{t:table2}
}
\begin{tabular}{lrr}
\hline
 & Model~1 & Model~2 \\
\hline
    $M / M_\odot$ &  3175 & 2037 \\ 
    $f_{\rm b}$ & 0.53 & 0.60 \\
    $T/T_{\rm rh}$ & 9 & 13 \\      
    $R_{\rm t}$/pc & 20.8 & 15.2 \\
    $R_{\rm h}$/pc & 4.9 & 3.8 \\
    $M_{\rm L} / M_\odot$ &  1987 & 1488 \\ 
    $M_{\rm L10} / M_\odot$ &  1730 & 1342 \\ 
    $R_{\rm h,L10}$/pc & 3.0 & 2.7 \\
\hline
\end{tabular}
\end{minipage}
\end{table*}

\begin{table*}
\begin{minipage}{126mm}
\caption{
Stellar population results at $4\,$Gyr for simulations performed 
in this work (see text for details). 
\label{t:table3}
}
\begin{tabular}{lrr}
\hline
 & Model~1 & Model~2 \\
\hline
    $N_{\rm BS}$ & 1 & 20 \\
    $N_{\rm BS,bin}$ & 1 & 9 \\
    $N_{\rm BS} / N_{\rm MS,2to}$ & 0.01 & 0.18 \\
    $R_{\rm h,BS}$/pc & -- & 1.1 \\
    $N_{\rm RS}$ & 2 & 6 \\
    $N_{\rm CV}$ & 3 & 1 \\
    $f_{\rm WD}$ & 0.16 & 0.15 \\
    $R_{\rm h,WD}$/pc & 0.3 & 0.6 \\
\hline
\end{tabular}
\end{minipage}
\end{table*}

\begin{table*}
\begin{minipage}{126mm}
\caption{
Details of the blue stragglers observed in Model~2 at $4\,$Gyr. 
Columns are the star ID number for the BS, mass of the BS, 
its $V$-band magnitude and $(B-V)$ colour (that of the BS or unresolved BS-binary), 
the radial position in the cluster at $4\,$Gyr, and 
the time at which the BS obtained its current mass. 
If the BS is in a binary the companion type is given in Column~7 followed by 
the companion mass, orbital period and eccentricity. 
The final column gives a classification of the evolution history of the BS using 
the following key: 
prim = primordial binary; A = Case~A mass transfer 
leading to coalescence; B = Case~B mass transfer; C = Case~C mass transfer; 
coll = collision in eccentric binary; 
exch = exchange interaction; 
and pert = perturbation to orbit.  
\label{t:table4}
}
\begin{tabular}{rcccrrlcrcl}
\hline
ID\# & $M/M_\odot$ & $V$ & $(B-V)$ & $r / {\rm pc}$ & $T_0 / {\rm Myr}$ & 
type & $M_2/M_\odot$ & $P / {\rm d}$ & e & history \\
\hline
 3289 & 2.10 & 10.95 & 0.12 &   1.03 & 3613 & MS & 1.3 & 81.3 & 0.86 & exch--coll--coll \\
 1418 & 2.09 & 10.82 & 0.20 &   0.37 & 3844 & giant & 0.7 & 1.95 & 0.00 & C--exch--B \\
 2203 & 2.08 & 10.94 & 0.17 &   1.10 & 3480 & MS & 0.8 & 363 & 0.30 & pert--coll--exch--pert \\
 2411 & 1.97 & 11.61 & 0.07 &   0.49 & 3657 & -- & -- & -- & -- & prim--pert--A \\ 
 2565 & 1.89 & 11.44 & 0.22 & 19.38 & 3652 & -- & -- & -- & -- & prim--pert--coll \\ 
 1613 & 1.88 & 11.36 & 0.34 &   0.92 & 2871 & MS & 0.9 & 11749 & 0.27 & exch--A--exch \\
 2321 & 1.88 & 11.71 & 0.13 &   1.74 & 3971 & MS & 0.7 & 26.9 & 0.65 & prim--pert--coll \\
 2737 & 1.80 & 11.56 & 0.39 &   0.91 & 2549 & -- & -- & -- & -- & prim--A \\ 
 2855 & 1.74 & 11.84 & 0.28 &   2.11 & 2946 & -- & -- & -- & -- & prim--A \\ 
 3835 & 1.73 & 11.90 & 0.26 &   0.31 & 3798 & MS & 1.0 & 19498 & 0.69 & pert--coll--exch--coll \\
 2973 & 1.69 & 11.93 & 0.38 &   1.10 & 3115 & -- & -- & -- & -- & prim--pert--coll \\ 
 3021 & 1.67 & 12.17 & 0.27 &   2.96 & 3313 & -- & -- & -- & -- & prim--pert--A \\ 
 3157 & 1.63 & 12.04 & 0.56 &   3.08 & 1948 & -- & -- & -- & -- & prim--A \\ 
 3121 & 1.64 & 12.32 & 0.27 &   0.66 & 3885 & MS & 0.3 & 28184 & 0.72 & pert--coll--exch \\
 3207 & 1.61 & 12.27 & 0.34 &   2.44 & 3803 & -- & -- & -- & -- & prim--A \\ 
 3445 & 1.53 & 12.36 & 0.60 &   4.42 & 2768 & MS & 0.3 & 8511 & 0.58 & pert--coll--exch \\ 
 3523 & 1.51 & 12.77 & 0.35 &   0.75 & 3896 & -- & -- & -- & -- & prim--A \\ 
 3877 & 1.40 & 12.82 & 0.52 &   1.31 & 1425 & -- & -- & -- & -- & prim--A \\ 
 3885 & 1.40 & 12.80 & 0.54 &   1.65 & 1241 & -- & -- & -- & -- & prim--A \\ 
 1378 & 1.36 & 12.92 & 0.57 &   1.55 & 1957 & WD & 0.6 & 1660 & 0.25 & prim--C--pert \\
\hline
\end{tabular}
\end{minipage}
\end{table*}

\begin{table*}
\hspace*{-4.0cm}
\begin{minipage}{126mm}
\caption{
Detailed description of the formation scenario for some of the BSs  
listed in Table~\ref{t:table4}. 
\label{t:table5}
}
\begin{tabular}{rl}
\hline
ID\# & explanation \\
\hline
 1378 & This primordial binary began with an orbital period of $14454\,$d, 
                an eccentricity of $0.83$ and stellar masses of 1.75 and $1.21 M_\odot$. \\ 
            & After $1954\,$Myr the orbit began to circularize 
                with the more massive star on the AGB. The primary filled its Roche lobe shortly \\ 
            & afterwards when the orbit was circular with $P = 2\,203\,$d. 
                At this point the masses were $1.02$ and $1.23 M_\odot$ so that CE evolution \\ 
            & was avoided and stable Case~C mass-transfer began. 
                This phase ended when the primary had shed its entire envelope to become \\ 
            & a $0.62 M_\odot$ CO WD in a circular binary of period $1\,819\,$d with a 
                $1.36 M_\odot$ MS star companion. This was perturbed by a third star \\ 
             & at $T = 2\,504\,$Myr 
                when $0.35\,$pc from the cluster centre. The period was reduced to 
                $1\,659\,$d and an eccentricity of 0.25 was induced. \\  \hline 
 1418 & The proto-BS originated as a $1.23 M_\odot$ star in a binary of 
                period $4\,466\,$d and eccentricity 0.7 with a $1.67 M_\odot$ companion. \\ 
            & After $2\,243\,$Myr the initially more massive star had evolved to the 
                AGB and wind mass loss had reduced it to $1.21 M_\odot$ while \\ 
            &  tidal forces had circularized the orbit ($P = 1\,750\,$d). 
                 At this point Case~C mass-transfer began. This ended with a $0.64 M_\odot$ \\ 
            & CO WD and a $1.45 M_\odot$ MS star in a circular orbit  
                with $P = 1\,445\,$d. At $T = 3\,243\,$Myr it was involved in an exchange with \\ 
             & a $1.35 M_\odot$ star. The WD was ejected to leave a binary 
                 with $e = 0.98$ and $P = 3\,631\,$d. At $T = 3\,844\,$Myr the $1.35 M_\odot$ star \\ 
              & evolved off the MS, tides circularized the orbit 
                 and Case~B mass-transfer started (ongoing at $4\,$Gyr). \\ \hline 
 1613 & A primordial binary of period $47\,860\,$d and eccentricity 0.8 containing stars 
               of mass $1.46$ and $0.3 M_\odot$ became involved \\ 
            & in a four-body interaction with another primordial binary 
               ($P = 3\,782\,$d, $e = 0.65$ and masses of $0.42$ and $0.38 M_\odot$) after \\ 
            & $2\,089\,$Myr. From this a short-period eccentric binary containing the 
                $1.46$ and $0.42 M_\odot$ stars was formed ($P = 1.1\,$d, \\ 
             & $e = 0.8$). At $T = 2\,871\,$Myr the binary had circularized and the $1.46 M_\odot$ MS star 
                 began Case~A mass transfer quickly  \\ 
             & followed by coalescence. The $1.88 M_\odot$ single BS became involved 
                 in a three-body hierarchy at $T = 2\,932\,$Myr with a wide  \\ 
              &  primordial binary ($0.86$ and $0.81 M_\odot$) in the cluster core. Eventually the least 
                  massive star was ejected from the system. \\ \hline 
 2203 & The proto-BS began life as a $1.23 M_\odot$ star with a $0.85 M_\odot$ 
                companion in a primordial binary of period $363\,$d and an \\ 
            & eccentricity of 0.3. After $3\,479\,$Myr a binary--binary encounter left this 
                binary as the inner component of a four-body \\ 
            & system with stars of mass $0.83$ and $0.58 M_\odot$. 
                The eccentricity of the inner binary was driven up to 0.99 so that the  \\ 
             & stars collided and formed the $2.08 M_\odot$ BS. The $0.58 M_\odot$ star  
                  escaped the system and the BS remained bound to the \\ 
              & $0.83 M_\odot$ MS star 
                  ($P = 372\,$d, $e = 0.84$). A subsequent 
                interaction reduced the period to $363\,$d and the eccentricity to 0.3.   \\ \hline  
 2321 & This primordial binary was originally comprised of $1.29$ and $0.73 M_\odot$ 
                stars in a circular orbit with a period of $31\,$d. \\ 
            & At $T = 3\,360\,$Myr while in the core of the cluster this binary had a close 
                encounter with another primordial binary to  \\ 
            & form a quadruplet system. An eccentricity of 0.15 had been induced into the binary 
                at this point.  At $T = 3\,971\,$Myr \\ 
             & the $1.29 M_\odot$ star collided with a $0.59 M_\odot$ 
                MS star -- the relative eccentricity of these two stars had reached 0.99 -- to \\  
             & form the $1.88 M_\odot$ BS. 
                The BS remained bound to its original companion
                 and the fourth star ($0.41 M_\odot$) was ejected. \\ \hline 
 2411 & This began in a circular primordial binary with $1\,$d period and component masses 
                $1.29$ and $0.68 M_\odot$. Case~A mass-transfer  \\ 
            & was expected to start at $4\,320\,$Myr but perturbation at $3\,100\,$Myr while binary 
                was in the core hardened the orbit and \\ 
             & mass-transfer began at $3\,260\,$Myr. Angular momentum loss 
                from the binary lead to coalescence of the stars at $3\,657\,$Myr.  \\ \hline 
 2565 & This started in a primordial binary with $3\,236\,$d period and eccentricity of 0.32. 
               The orbit was wide enough that interaction  \\ 
           & between the $0.95$ and $0.94 M_\odot$ stars was not expected. 
               Perturbation to the orbit while the binary was $1.0\,$pc from the cluster \\ 
           & centre increased the eccentricity to 0.99. 
               The orbit became chaotic and the stars collided and merged at $T = 3\,652\,$Myr. \\ \hline
 2973 & This is similar to $\#2565$. Interaction was not expected in a wide primordial binary 
               with $P = 4\,075\,$d and $e = 0.59$. A \\ 
           & perturbation pumped the eccentricity to 0.99 so that the $0.89$ and $0.80 M_\odot$ MS stars 
               collided and merged at $T = 3\,115\,$Myr. \\ \hline 
 3021 & This originated in a circular primordial binary with $P = 0.7\,$d and component masses 
                $1.01$ and $0.66 M_\odot$ that was \\ 
           &  expected to begin Case~A mass-transfer after $1\,413\,$Myr. Prior to this, at $T = 1\,010\,$Myr, 
                the binary was involved in a \\ 
           & short-lived exchange encounter. The primordial binary emerged intact but the period 
               had increased enough to delay \\ 
           & the onset of mass-transfer until $T = 2\,900\,$Myr. The stars merged at 
               $T = 3\,313\,$Myr. \\ \hline 
 3121 & This was a  primordial binary composed of $1.09$ and $0.54 M_\odot$ stars with 
                a period of $562\,$d and an eccentricity of $0.31$ which \\ 
            & became part of a sextuplet after $3\,762\,$Myr. The eccentricity of the binary was 
                increased by perturbations from the \\ 
             & other members until at $T = 3\,885\,$Myr the orbit 
                became chaotic, the eccentricity reached 0.99 and the two stars collided. \\ 
              & The resulting $1.64 M_\odot$ BS remained bound to a $0.33 M_\odot$ 
                member of the sextuplet to give the binary observed at $4\,$Gyr. \\ \hline 
 3289 & Two short-period circular primordial binaries 
               ($P = 1.3\,$d, $M_1 = 0.82 M_\odot$, $M_2 = 0.76 M_\odot$ and  
                $P = 4.2\,$d, $M_1 = 1.25 M_\odot$, \\ 
             & $M_2 = 0.51 M_\odot$) became embroiled in a four-body system after $3\,611\,$Myr. 
                The $0.82$ and $0.51 M_\odot$ stars formed an inner \\ 
             & binary and collided after the eccentricity was pumped up to unity. The collision product then 
               collided with the $0.76 M_\odot$ \\ 
             & star to form a $2.09 M_\odot$ BS. 
               The BS remained bound to the $M_1 = 1.25 M_\odot$ MS star in an 
               eccentric orbit with $P = 81\,$d. \\ \hline 
 3445 & The proto-BS originated as a $0.78 M_\odot$ star in a $2.9\,$d circular orbit 
                with a $0.75 M_\odot$ companion. No interaction between \\ 
             & the stars was expected in this short-period primordial binary. After $2\,768\,$Myr the 
                binary formed a triple system with \\ 
             & a $0.25 M_\odot$ star. The close presence of the third 
                star induced an eccentricity of 0.84 into the binary orbit so that the \\ 
             & stars came into contact and merged. The $1.53 M_\odot$ merged star 
                remained bound to the $0.25 M_\odot$ star. \\ \hline 
 3835 & A primordial binary consisting of $0.82$ and $0.60 M_\odot$ stars in a 
               $1.8\,$d circular orbit became involved in a four-body system with \\  
            & another binary (formed earlier from an exchange) at $T = 3\,797\,$Myr. 
               The stars in the primordial binary collided and merged   \\ 
            & and this new star then collided with a $0.31 M_\odot$ star 
                to form a $1.73 M_\odot$ BS. The BS remained bound to the fourth member of  \\ 
            & the system, a $1.04 M_\odot$ 
               MS star, in a long-period eccentric orbit. \\
\hline
\end{tabular}
\end{minipage}
\end{table*}

\begin{table*}
\begin{minipage}{126mm}
\caption{
Details of the RS CVn binaries observed in Model~2 at $4\,$Gyr. 
The first column gives the star ID number for the subgiant star 
and this is followed by the mass of the subgiant. 
Column~3 gives the mass of the MS star companion. 
The $V$-band magnitude and $(B-V)$ colour of the unresolved binary 
are then given. 
The sixth column gives the radial position in the cluster of the binary 
at $4\,$Gyr. 
The orbital period and eccentricity are given in Columns~7 and 8, 
respectively. 
Some remarks on each binary are provided in the final column. 
\label{t:table6}
}
\begin{tabular}{rcccccrcl}
\hline
ID\# & $M_1/M_\odot$ & $M_2/M_\odot$ & $V$ & $(B-V)$ & $r / {\rm pc}$ & 
$P / {\rm d}$ & e & remarks \\
\hline
 1568 & 1.33 & 0.95 & 12.50 & 0.60 & 0.27 & 1.6 & 0.0 & exchange \\
 1799 & 1.36 & 1.08 & 12.30 & 0.88 & 4.91 & 5.0 & 0.0 & primordial \\
 2335 & 1.33 & 0.68 & 12.69 & 0.68 & 0.28 & 3.0 & 0.0 & primordial \\
 2383 & 1.37 & 0.62 & 10.49 & 1.17 & 1.32 & 19.8 & 0.0 & primordial \\
 2633 & 0.63 & 1.19 & 13.82 & 0.64 & 1.11 & 0.4 & 0.0 & primordial; semi-detached \\
 2873 & 1.33 & 0.40 & 12.05 & 0.63 & 8.46 & 6.2 & 0.0 & primordial \\
\hline
\end{tabular}
\end{minipage}
\end{table*}

\label{lastpage}


\newpage 

\begin{thebibliography}{}
\bibitem[\protect\citeauthoryear{Aarseth}{1966}]{aar66} Aarseth S. J., 1966, 
    in Kontopoulos G., ed, 
    Proc, IAU Symp. 25, 
    The Theory of Orbits in the Solar System and in Stellar System. 
    Academic Press, London, p. 141 
\bibitem[\protect\citeauthoryear{Aarseth}{1996}]{aar96} Aarseth S. J., 1996, 
    in Milone E., Mermilliod J.-C., eds, 
    ASP Conf. Series, Vol. 90, 
    The Origins, Evolution, and Destinies of Binary Stars in Clusters. 
    ASP, San Francisco, p. 423 
\bibitem[\protect\citeauthoryear{Aarseth}{1999}]{aar99} Aarseth S. J.,1999, PASP, 111, 1333
\bibitem[\protect\citeauthoryear{Aarseth}{2003}]{aar03} Aarseth S.J., 2003, 
    Gravitational N-body Simulations: Tools and Algorithms 
    (Cambridge Monographs on Mathematical Physics). 
    Cambridge University Press, Cambridge
\bibitem[\protect\citeauthoryear{Aarseth, H\'{e}non \& Wielen}{1974}]{aar74} Aarseth S.,
    H\'{e}non M., Wielen R., 1974, A\&A, 37, 183 
\bibitem[\protect\citeauthoryear{Abt}{1983}]{abt83} Abt H.A., 1983, ARA\&A, 21, 343 
\bibitem[\protect\citeauthoryear{Ahumada \& Lapasset}{1995}]{ahu95} Ahumada J., 
    Lapasset E., 1995, A\&AS, 109, 375 
\bibitem[\protect\citeauthoryear{Baumgardt}{1998}]{bau98} Baumgardt H., 1998, A\&A, 330, 480 
\bibitem[\protect\citeauthoryear{Baumgardt}{2001}]{bau01} Baumgardt H., 2001, MNRAS, 325, 1323
\bibitem[\protect\citeauthoryear{Baumgardt \& Makino}{2003}]{bau03} Baumgardt H., 
    Makino J., 2003, MNRAS, 340, 227
\bibitem[\protect\citeauthoryear{Belloni, Verbunt \& Mathieu}{1998}]{bel98} Belloni T., Verbunt F., 
    Mathieu R.D., 1998, A\&A, 339, 431 
\bibitem[\protect\citeauthoryear{Bergeron, Wesemael \& Beauchamp}{1995}]{ber95} Bergeron P., 
    Wesemael F., Beauchamp A., 1995, PASP, 107, 1047
\bibitem[\protect\citeauthoryear{Bonatto \& Bica}{2003}]{bon03} Bonatto Ch., Bica E., 2003, 
    A\&A, 405, 525
\bibitem[\protect\citeauthoryear{Bonatto \& Bica}{2005}]{bon05} Bonatto Ch., Bica E., 2005, 
    astro-ph/0503589 
\bibitem[\protect\citeauthoryear{Carraro \& Chiosi}{1994}]{car94} Carraro G., Chiosi C.,  
    1994, A\&A, 288, 751
\bibitem[\protect\citeauthoryear{Casertano \& Hut}{1985}]{cas85} Casertano S., Hut P., 
    1985, ApJ, 298, 80 
\bibitem[\protect\citeauthoryear{Chernoff \& Weinberg}{1990}]{che90} Chernoff D.F., Weinberg M.D.,  
    1990, ApJ, 351, 121
\bibitem[\protect\citeauthoryear{Davies, Piotto \& De Angeli}{2004}]{dav04} Davies M.B., 
    Piotto G., De Angeli F., 2004, MNRAS, 349, 129
\bibitem[\protect\citeauthoryear{Djorgovski}{1988}]{djo88} Djorgovski S., 1988, 
    in Proc. IAU Symp. 126, 
    The Harlow-Shapley Symposium on Globular Cluster Systems in Galaxies. 
    Kluwer, Dordrecht, p. 333
\bibitem[\protect\citeauthoryear{Duquennoy \& Mayor}{1991}]{duq91} Duquennoy A., Mayor M., 
    1991, A\&A, 248, 485
\bibitem[\protect\citeauthoryear{Eggleton, Fitchett \& Tout}{1989}]{egg89} Eggleton P.P., 
    Fitchett M., Tout C.A., 1989, ApJ, 347, 998 
\bibitem[\protect\citeauthoryear{Elson, Fall \& Freeman}{1987}]{els87} Elson R.A.W., 
    Fall S.M., Freeman K.C., 1987, ApJ, 323, 54 
\bibitem[\protect\citeauthoryear{Fan et al.}{1996}]{fan96} Fan X., et al., 1996, AJ, 112, 628
\bibitem[\protect\citeauthoryear{Giersz \& Heggie}{1994}]{gie94} Giersz M., Heggie D.C., 1994, 
    MNRAS, 268, 257 
\bibitem[\protect\citeauthoryear{Girard et al.}{1989}]{gir89} Girard T.M., Grundy W.M., 
     L\'{o}pez C.E., van Altena W.F., 1989, AJ, 98, 227 
\bibitem[\protect\citeauthoryear{Grindlay et al.}{2001}]{gri01} Grindlay J.E., Heinke C., Edmonds P.D., 
    Murray S.S., 2001, Science, 292, 2290
\bibitem[\protect\citeauthoryear{Hall}{1976}]{hal76} Hall D.S., 1976, 
    in Fitch W.S., ed, 
    Proc. IAU Colloq. 29, 
    Multiple Periodic Variable Stars. Reidel, Dordrecht, p. 287 
\bibitem[\protect\citeauthoryear{Heggie}{1975}]{heg75} Heggie D.C., 1975, MNRAS, 173, 729
\bibitem[\protect\citeauthoryear{Heggie, Hut \& McMillan}{1996}]{heg96} Heggie D.C., Hut P., 
    McMillan S.L.W., 1996, ApJ, 467, 359 
\bibitem[\protect\citeauthoryear{Hobbs \& Thorburn}{1991}]{hob91} Hobbs L.M., Thorburn J.A., 
    1991, AJ, 102, 1070 
\bibitem[\protect\citeauthoryear{Hurley \& Shara}{2003}]{hur03} Hurley J. R., 
    Shara M.M., 2003, ApJ, 589, 179 
\bibitem[\protect\citeauthoryear{Hurley, Pols \& Tout}{2000}]{hur00} Hurley J. R., Pols O.R., 
    Tout C. A., 2000, MNRAS, 315, 543
\bibitem[\protect\citeauthoryear{Hurley et al.}{2001}]{hur01} Hurley J. R., Tout C. A., 
    Aarseth S. J., Pols,O.R., 2001, MNRAS, 323, 630
\bibitem[\protect\citeauthoryear{Hurley et al.}{2004}]{hur04} Hurley J. R., Tout C. A., 
    Aarseth S. J., Pols,O.R., 2004, MNRAS, 355, 1207 
\bibitem[\protect\citeauthoryear{Hurley, Tout \& Pols}{2002}]{hur02} Hurley J. R., Tout C. A., 
    Pols,O.R., 2002, MNRAS, 329, 897
\bibitem[\protect\citeauthoryear{Iben}{1990}]{ibe90} Iben I.Jr., 1990, ApJ, 353, 215 
\bibitem[\protect\citeauthoryear{Janes \& Phelps}{1994}]{jan94} Janes K.A., Phelps R.L., 
    1994, AJ, 108, 1773
\bibitem[\protect\citeauthoryear{Kafka et al.}{2004}]{kaf04} Kafka S., Gibbs D.G.II., Henden A.A., 
    Honeycutt R.K., 2004, AJ, 127, 1622 
\bibitem[\protect\citeauthoryear{Kalirai et al.}{2001}]{kal01} Kalirai J. S., et al., 2001, AJ, 122, 266
\bibitem[\protect\citeauthoryear{Kalirai et al.}{2003}]{kal03} Kalirai J. S., Fahlman G.G., 
    Richer H.B., Ventura P., 2003, AJ, 126, 1402
\bibitem[\protect\citeauthoryear{King}{1962}]{kin62} King I.R., 1962, AJ, 67, 471
\bibitem[\protect\citeauthoryear{King}{1966}]{kin66} King I.R., 1966, AJ, 71, 64  
\bibitem[\protect\citeauthoryear{Kippenhahn, Wiegert  \& Hoffmeister}{1967}]{kip67} Kippenhahn R., 
    Wiegert A., Hoffmeister R., 1967, 
    in Alder B., Fernbach S., Rotenburg M., eds, 
    Methods in Computational Physics, Vol. 7, 
    Astrophysics, New York, p. 129 
\bibitem[\protect\citeauthoryear{Kochanek}{1992}]{koc92} Kochanel C.S., 1992, ApJ, 385, 604
\bibitem[\protect\citeauthoryear{Kozai}{1962}]{koz62} Kozai Y., 1962, AJ, 67, 591
\bibitem[\protect\citeauthoryear{Kroupa}{1995a}]{kro95a} Kroupa P., 1995a, MNRAS, 277, 1491 
\bibitem[\protect\citeauthoryear{Kroupa}{1995b}]{kro95b} Kroupa P., 1995b, MNRAS, 277, 1507
\bibitem[\protect\citeauthoryear{Kroupa, Aarseth \& Hurley}{2001}]{kro01} Kroupa P., Aarseth S., 
    Hurley J., 2001, MNRAS, 321, 699
\bibitem[\protect\citeauthoryear{Kroupa, Tout \& Gilmore}{1991}]{kro91} Kroupa P., Tout C. A., 
    Gilmore G., 1991, MNRAS, 251, 293
\bibitem[\protect\citeauthoryear{Kroupa, Tout \& Gilmore}{1993}]{kro93} Kroupa P., Tout C. A., 
    Gilmore G., 1993, MNRAS, 262, 545
\bibitem[\protect\citeauthoryear{Kurucz}{1992}]{kur92} Kurucz R.L., 1992, 
    in Barbuy B., Renzini A., eds, 
    Proc. IAU Symp. 149, 
    The Stellar Populations of Galaxies. 
    Kluwer, Dordrecht, p. 225
\bibitem[\protect\citeauthoryear{Lang}{1992}]{lan92} Lang K.R., 1992, Astrophysical Data. 
    Springer-Verlag, Berlin 
\bibitem[\protect\citeauthoryear{Latham \& Milone}{1996}]{lat96} Latham D.W., Milone A.A.E., 1996, 
    in Milone E.F., Mermilliod J.-C., eds, 
     ASP Conf. Series, Vol. 90, 
    The Origins, Evolution, and Destinies of Binary Stars in Clusters. 
    ASP, San Francisco, p. 385
\bibitem[\protect\citeauthoryear{Leonard}{1996}]{leo96} Leonard P.J.T., 1996, ApJ, 470, 521
\bibitem[\protect\citeauthoryear{Lombardi, Rasio \& Shapiro}{1996}]{lom96} Lombardi J.C., 
    Rasio F.A., Shapiro S.L., 1996, ApJ, 468, 797 
\bibitem[\protect\citeauthoryear{Lombardi et al.}{2003}]{lom03} Lombardi J.C., 
    Thrall A.P., Deneva J.S., Fleming S.W., Grabowski P.E., 2003, MNRAS, 345, 762 
\bibitem[\protect\citeauthoryear{Mackey \& Gilmore}{2003}]{mac03} Mackey A.D., Gilmore G.F.,  
    2003, MNRAS, 338, 85 
\bibitem[\protect\citeauthoryear{Makino}{1991}]{mak91} Makino J., 1991, ApJ, 369, 200 
\bibitem[\protect\citeauthoryear{Makino}{2002}]{mak02} Makino J., 2002, 
    in Shara M.M., ed, 
    ASP Conference Series 263, 
    Stellar Collisions, Mergers and their Consequences. 
    ASP, San Francisco, p. 389
\bibitem[\protect\citeauthoryear{Makino \& Taiji}{1998}]{mak98} Makino J., Taiji M., 1998, 
    Scientific Simulations with Special-Purpose Computers -- the GRAPE Systems. 
    Wiley, New York 
\bibitem[\protect\citeauthoryear{Makino et al.}{2003}]{mak03} Makino J., Fukushige T., 
    Koga M., Namura K., 2003, PASJ, 55, 1163 
\bibitem[\protect\citeauthoryear{Mathieu et al.}{2003}]{mat03} Mathieu R.D., van den Berg M., 
    Torres G., Latham D., Verbunt F., Stassun K., 2003, AJ, 125, 246 
\bibitem[\protect\citeauthoryear{Mapelli et al.}{2004}]{map04} Mapelli M., Sigurdsson S., 
    Colpi M., Ferraro F.R., Possenti A., Rood R.T., Sills A., Beccari G., 2004, ApJ, 605, L29 
\bibitem[\protect\citeauthoryear{Mardling \& Aarseth}{2001}]{mar01} Mardling R.A., 
    Aarseth S.J., 2001, MNRAS, 321, 398
\bibitem[\protect\citeauthoryear{McMillan}{1986}]{mcm86} McMillan S.L.W., 1986, 
    in Hut P., McMillan S.L.W., eds, 
    The Use of Supercomputers in Stellar Dynamics. 
    Springer-Verlag, Berlin, p. 156
\bibitem[\protect\citeauthoryear{Mikkola \& Aarseth}{1993}]{mik93} Mikkola S., Aarseth S.J., 
    1993, Celest. Mech. Dyn. Astron., 57, 439
\bibitem[\protect\citeauthoryear{Mikkola \& Aarseth}{1998}]{mik98} Mikkola S., Aarseth S.J., 
    1998, New Astronomy, 3, 309 
\bibitem[\protect\citeauthoryear{Milone \& Latham}{1992}]{mil92} Milone A.A.E., Latham D.W., 1992, 
    in Kondo, Y., Sister\'{o} R.F., Polidan R.S., eds, 
    Proc. IAU Symp. 151, 
    Evolutionary Processes in Interacting Binary Stars. 
    Kluwer, Dordrecht, p. 475 
\bibitem[\protect\citeauthoryear{Montgomery, Marschall \& Janes}{1993}]{mon93} 
    Montgomery K.A., Marschall L.A., Janes K.A., 1993, AJ, 106, 181
\bibitem[\protect\citeauthoryear{Nomoto \& Kondo}{1991}]{nom91} Nomoto K., Kondo Y., 
    1991, ApJ, 367, L19 
\bibitem[\protect\citeauthoryear{Piotto et al.}{2002}]{pio02} Piotto G., et al., 2002, A\&A, 391, 945 
\bibitem[\protect\citeauthoryear{Plummer}{1911}]{plu11} Plummer H.C., 1911, MNRAS, 71, 460 
\bibitem[\protect\citeauthoryear{Pols et al.}{2003}]{pol03} Pols O.R., Karakas A.I., 
    Lattanzio J.C., Tout C.A., 2003, ASPC, 303, 290 
\bibitem[\protect\citeauthoryear{Pols et al.}{1998}]{pol98} Pols O.R., Schr\"{o}der K.-P., Hurley J.R., 
    Tout C.A., Eggleton P.P., 1998, MNRAS, 298, 525
\bibitem[\protect\citeauthoryear{Pooley et al.}{2003}]{poo03} Pooley D., et al., 2003, ApJ, 591, L131
\bibitem[\protect\citeauthoryear{Portegies Zwart et al.}{2001}]{por01} Portegies Zwart S.F., 
    McMillan S.L.W., Hut P., Makino J., 2001, MNRAS, 321, 199
\bibitem[\protect\citeauthoryear{Portegies Zwart et al.}{2004}]{por04} Portegies Zwart S.F., 
    Hut P., McMillan S.L.W., Makino J., 2004, MNRAS, 351, 473 
\bibitem[\protect\citeauthoryear{Press \& Teukolsky}{1977}]{pre77} Press W.H., Teukolsky S.A., 
    1977, ApJ, 213, 183 
\bibitem[\protect\citeauthoryear{Preston \& Sneden}{2000}]{pre00} Preston G.W., Sneden C., 
    2000, AJ, 120, 1014 
\bibitem[\protect\citeauthoryear{Randich}{1997}]{ran97} Randich S., 1997, MmSAI, 68, 971
\bibitem[\protect\citeauthoryear{Richer et al.}{1998}]{ric98} Richer H.B., Fahlman G.G., 
    Rosvick J., Ibata R., 1998, ApJ, 116, L91
\bibitem[\protect\citeauthoryear{Richer et al.}{2004}]{ric04} Richer H.B., et al., 2004, AJ, 127, 2771 
\bibitem[\protect\citeauthoryear{Saio \& Nomoto}{1998}]{sai98} Saio H., Nomot K., 1998, ApJ, 500, 388
\bibitem[\protect\citeauthoryear{Sandquist}{2004}]{san04} Sandquist E.L., 2004, MNRAS, 347, 101
\bibitem[\protect\citeauthoryear{Sandquist \& Shetrone}{2003}]{san03} Sandquist E.L., 
    Shetrone M.D., 2003, AJ, 125, 2173 
\bibitem[\protect\citeauthoryear{Schertl et al.}{2003}]{sch03} Schertl D., Balega Y.Y., 
    Preibisch Th., Weigelt G., 2003, A\&A, 402, 267 
\bibitem[\protect\citeauthoryear{Shara \& Hurley}{2002}]{sha02} Shara M.M., 
    Hurley J.R., 2002, ApJ, 571, 830
\bibitem[\protect\citeauthoryear{Sills et al.}{2001}]{sil01} Sills A., Faber J.A., Lombardi J.C., 
    Rasio F.A., Warren A.R., 2001, ApJ, 548, 323
\bibitem[\protect\citeauthoryear{Sills et al.}{2003}]{sil03} Sills A., et al., 2003, New Astronomy, 8, 605 
\bibitem[\protect\citeauthoryear{Stetson, McClure \& VandenBerg}{2004}]{ste04} Stetson P.B., 
    McClure R.D., VandenBerg D.A., 2004, PASP, 116, 1012 
\bibitem[\protect\citeauthoryear{Tautvai\u{s}ien\.{e} et al.}{2000}]{tau00} Tautvai\u{s}ien\.{e} G., 
    Edvardsson B., Tuominen I., Ilyin, I., 2000, A\&A, 360, 499
\bibitem[\protect\citeauthoryear{Terlevich}{1987}]{ter87} Terlevich E., 1987, MNRAS, 224, 193
\bibitem[\protect\citeauthoryear{Tokovinin}{1997}]{tok97} Tokovinin A.A., 1997, A\&AS, 124, 75
\bibitem[\protect\citeauthoryear{van Albada}{1968}]{van68} van Albada T.S., 1968, 
    Bull. Astron. Inst. Neth., 19, 479 
\bibitem[\protect\citeauthoryear{VandenBerg \& Stetson}{2004}]{van04} VandenBerg D.A., 
    Stetson P.B., 2004, PASP, 116, 997
\bibitem[\protect\citeauthoryear{van den Berg, Verbunt \& Mathieu}{1999}]{van99} van den Berg M., 
    Verbunt F., Mathieu R.D., 1999, A\&A, 347, 866
\bibitem[\protect\citeauthoryear{van den Berg et al.}{2001}]{van01} van den Berg M., Orosz J., 
    Verbunt F., Stassun K., 2001, A\&A, 375, 375
\bibitem[\protect\citeauthoryear{van den Berg et al.}{2004}]{van04} van den Berg M., Tagliaferri G., 
    Belloni T., Verbunt F., 2004, A\&A, 418, 509 
\bibitem[\protect\citeauthoryear{Vesperini}{1997}]{ves97} Vesperini E., 1997, MNRAS, 287, 915
\bibitem[\protect\citeauthoryear{von Hoerner}{1960}]{von60} von Hoerner S., 1960, 
    Z. Astrophys., 50, 184 
\bibitem[\protect\citeauthoryear{Walter}{1982}]{wal82} Walter F.M., 1982, ApJ, 253, 745 
\bibitem[\protect\citeauthoryear{Webbink}{1984}]{web84} Webbink R.F.,1984, ApJ, 277, 355 
\bibitem[\protect\citeauthoryear{Webbink}{1988}]{web88} Webbink R.F.,1998, 
    in Mikolajewska J., Friedjung M., Kenyon S.J., Viotto R., eds, 
    Proc. IAU Colloq. 103, 
   The Symbiotic Phenomenon. Kluwer, Dordrecht, p. 311
\bibitem[\protect\citeauthoryear{Wilkinson et al.}{2003}]{wil03} Wilkinson M.I., Hurley J.R., 
   Mackey A.D., Gilmore G.F., Tout C.A., 2003, MNRAS, 343, 1025  
\bibitem[\protect\citeauthoryear{Yungelson \& Livio}{2000}]{yun00} Yungelson L., Livio M., 
    2000, ApJ, 528, 108 
\end{thebibliography}
\end{document}